\documentclass[a4paper,12pt]{article}

\usepackage{amsmath}
\usepackage{amssymb}
\usepackage{amsfonts}
\usepackage{amscd}
\usepackage{amsthm}
\usepackage{latexsym}
\usepackage{amsbsy}
\usepackage{bbm}
\usepackage[english]{babel}
\usepackage{graphicx}
\usepackage{psfrag}
\usepackage[breaklinks=true]{hyperref}

\hypersetup{
     colorlinks=true,
     citecolor=blue,
     linkcolor=blue,
     urlcolor=black
   }

\begin{document}

\begin{center}

\vspace{.7cm}
{\LARGE\bf Holographic Renormalization for $z=2$\\\vspace{.2cm} Lifshitz Space-Times from AdS} \\

\vspace{1.2cm}

{{\bf Wissam Chemissany$^a$, David Geissb\"uhler$^b$, Jelle Hartong$^c$, Blaise Rollier$^b$}\\
\vspace{1cm}
 $^a$ {\small\slshape Department of Physics and Astronomy, University of Waterloo,\\ 
 Waterloo, Ontario, Canada, N2L 3G1}\\
 
 \vspace{0.2cm}

$^b${\small\slshape Albert Einstein Center for Fundamental Physics,\\
 Institute for Theoretical Physics,\\
 University of Bern,\\
 Sidlerstrasse 5, CH-3012 Bern,
 Switzerland}\\
 
\vspace{0.2cm}

$^c${\small\slshape The Niels Bohr Institute,\\
 University of Copenhagen,\\
 Blegdamsvej 17, DK-2100 Copenhagen \O,
 Denmark}}\\

\vspace{0.4cm}

{\small Email: \texttt{wchemiss@uwaterloo.ca, geissbuehler@itp.unibe.ch, hartong@nbi.ku.dk, rollier@itp.unibe.ch}}
\vspace{1cm}

 {\bf Abstract} \end{center} { }

Lifshitz space-times with critical exponent $z=2$ can be obtained by dimensional
reduction of Schroedinger space-times with critical exponent $z=0$. The latter
space-times are asymptotically AdS solutions of AdS gravity coupled to an
axion-dilaton system and can be uplifted to solutions of type IIB supergravity. This basic
observation is used to perform holographic renormalization for 4-dimensional asymptotically $z=2$ locally Lifshitz space-times by Scherk--Schwarz dimensional
reduction of the corresponding problem of holographic renormalization for
5-dimensional asymptotically locally AdS space-times coupled to an axion-dilaton
system. We can thus define and characterize a 4-dimensional asymptotically locally $z=2$ Lifshitz space-time in terms of 5-dimensional AdS boundary data.
In this setup the 4-dimensional structure of the Fefferman-Graham expansion and the structure of the counterterm action,
including the scale anomaly, will be discussed. We find that for asymptotically locally $z=2$ Lifshitz space-times obtained in this way there are two anomalies each with their own associated nonzero central charge. Both anomalies follow from the Scherk--Schwarz dimensional reduction of the 5-dimensional conformal anomaly of AdS gravity coupled to an axion-dilaton system. Together they make up an action that is of the Horava--Lifshitz type with nonzero potential term for $z=2$ conformal gravity.

\newpage

\begingroup
\hypersetup{linkcolor=black}
\tableofcontents
\endgroup

\section{Introduction}

Over the recent years we have witnessed a development in which it was realized that certain asymptotically AdS gravitational systems have features in common with systems encountered in the study of quantum phase transitions that occur in condensed matter physics when a system reaches a quantum critical point. See \cite{Hartnoll:2009sz,McGreevy:2009xe,Sachdev:2010ch} for some review papers. From the condensed matter point of view one is interested in the effective IR description of a system that in the UV consists of strongly coupled electrons. There exist cases where the effective field theory valid near the quantum critical point is described by a strongly coupled CFT \cite{Sachdev:2010ch}. The idea is to study such systems holographically by identifying sectors in holographically dual theories (consistent truncations of the complete theory) that via the concept of universality have the same universal properties as the condensed matter system one is interested in. On the gravity side this maps to a particular choice of matter fields on a background that becomes asymptotically AdS. 

However not all quantum critical points are described by CFTs. In general theories at a critical point are scale invariant with a scaling that is of the non-relativistic type:
\begin{equation*}
t\rightarrow\lambda^z t\qquad\&\qquad\vec x\rightarrow\lambda\vec x\,,
\end{equation*}
where $t$ and $\vec x$ are the respective time and space coordinates describing the system. The parameter $z$ is called the critical exponent. When $z\neq 1$ the theory can be either Lifshitz or Schr\"odinger invariant. Again such systems can occur at strong coupling. To study Lifshitz or Schr\"odinger invariant systems holographically we need to consider a space-time whose isometry group is the Lifshitz or the Schr\"odinger symmetry group. Such space-times are called Lifshitz \cite{Koroteev:2007yp,Kachru:2008yh} and Schr\"odinger \cite{Son:2008ye,Balasubramanian:2008dm} space-times, respectively.

Another interesting motivation to study (asymptotically) Lifshitz or Schr\"odinger space-times comes from the question: How general is holography? Since Lifshitz and Schr\"odinger space-times are no longer asymptotically AdS they form interesting examples to extend holographic techniques to asymptotically non-AdS space-times. 
In this work we will focus on space-times that are asymptotically locally $z=2$ Lifshitz\footnote{We briefly mention here that pure Lifshitz space-times suffer from IR singularities (divergent tidal forces in the bulk) \cite{Kachru:2008yh,Copsey:2010ya,Horowitz:2011gh}. In this project we will be primarily interested in the UV properties, i.e. close to the boundary, where there are no singularities.} in a sense to be made precise below (and agreeing with the definition given in \cite{Ross:2011gu}). For earlier work on asymptotically Lifshitz space-times and holographic renormalization see \cite{Ross:2009ar,Baggio:2011cp,Mann:2011hg,Ross:2011gu,Griffin:2011xs,Baggio:2011ha}. These studies have so far focussed on Lagrangians with no known string theory origin that contain gravity coupled to a massive vector field described by a Proca Lagrangian, but that do not contain dilatonic scalars. On the other hand we do know how to embed Lifshitz space-times into string theory \cite{Balasubramanian:2010uk,Donos:2010tu,Cassani:2011sv,Halmagyi:2011xh,Gregory:2010gx,Chemissany:2011mb}. Especially when $z=2$ the embedding of Lifshitz into string theory is quite straightforward. Here we will use the explicit model of \cite{Chemissany:2011mb} (based on \cite{Donos:2010tu,Cassani:2011sv}). This case is interesting for a number of reasons: 1). it is within the context of string theory, 2). there is an explicit relation with AdS via dimensional reduction (see below) and 3). it is explicitly $z=2$ which is a special value having properties that are different from generic $z$ values, so that it would be good to have an explicit detailed study of this case.

The basic idea of this paper is as follows. Lifshitz space-times with critical exponent $z=2$ can obtained by dimensional reduction of Schr\"odinger space-times with critical exponent $z=0$. The latter space-times are asymptotically AdS solutions of AdS gravity coupled to an axion-dilaton system. This basic observation is used to perform holographic renormalization for 4-dimensional asymptotically locally $z=2$ Lifshitz space-times by dimensional reduction of the corresponding problem of holographic renormalization for 5-dimensional asymptotically locally AdS space-times coupled to an axion-dilaton system.

Recently, interesting work appeared in relation to the Lifshitz scale anomaly \cite{Adam:2009gq,Gomes:2011di,Griffin:2011xs,Baggio:2011ha} generalizing the conformal anomaly for AdS gravity of \cite{Henningson:1998gx} to other values of $z$. In our setup we can make an explicit relation between the 5-dimensional AdS conformal anomaly (in the presence of an axion-dilaton system) and the 4-dimensional Lifshitz scale anomaly for $z=2$. We find that in the model we have studied there are two nonzero central charges and thus two associated anomalies for asymptotically locally $z=2$ Lifshitz space-times. In the remainder of this paper we will simply refer to that this AlLif space-times without explicitly writing $z=2$.

This paper is organized as follows. In section \ref{sec:HRaxiondilaton} we will review holographic renormalization for 5-dimensional AdS gravity coupled to an axion-dilaton system \cite{Papadimitriou:2011qb}. In the next section \ref{sec:SSreduction} we will work out the form of the 4-dimensional Fefferman--Graham expansions by Scherk--Schwarz reducing the Fefferman--Graham expansions of section \ref{sec:HRaxiondilaton}. Finally, in section \ref{sec:anomalies} we use these results to obtain the counterterm action of AlLif space-times by dimensional reduction of the counterterms of section \ref{sec:HRaxiondilaton} and we evaluate the anomaly counterterms on-shell using the results of section \ref{sec:SSreduction}.

\section{Holographic renormalization for AdS gravity coupled to an axion-dilaton field}\label{sec:HRaxiondilaton}

In this section we discuss the 5-dimensional model of AdS gravity coupled to an axion-dilaton system and review the holographic renormalization carried out in \cite{Papadimitriou:2011qb}. We will however not use the Hamiltonian formalism of \cite{Papadimitriou:2011qb}, but instead work within a Lagrangian framework. We will explicitly solve the equations of motion up to NNLO and discuss the local and anomaly counterterms as well as the one-point functions for asymptotically locally AdS (AlAdS) boundary conditions \cite{deHaro:2000xn,Papadimitriou:2005ii}.

\subsection{Fefferman--Graham expansions and counterterms}

The bulk action is
\begin{equation}\label{eq: Bulk lagrangian}
S_{\text{bulk}} =\frac{1}{2\kappa_5^2}\int_{\mathcal{M}}d^5x\mathcal{L}_{\text{bulk}}\,,
\end{equation}
where 
\begin{equation}
\mathcal{L}_{\text{bulk}}=\sqrt{-g}\left(R+12-\frac{1}{2}\partial_\mu\phi\partial^\mu\phi-\frac{1}{2}e^{2\phi}\partial_\mu\chi\partial^\mu\chi\right)\,.
\end{equation}
and where $\kappa_5^2=8\pi G_5$ with $G_5$ the 5-dimensional Newton's constant. The Gibbons--Hawking boundary action is given by
\begin{equation}\label{eq:GHaction}
S_{\text{GH}}=\frac{1}{\kappa_5^2}\int_{\partial\mathcal{M}}d^4x\sqrt{-h}K\,,
\end{equation}
where $h$ denotes the boundary metric. We have set the AdS$_5$ length equal to one. 

The equations of motion that we would like to obtain by varying $S_{\text{bulk}}+S_{\text{GH}}$ (supplied with additional boundary terms for asymptotically locally AdS boundary conditions) are
\begin{eqnarray}
\mathcal{E}_{\mu\nu} & = & G_{\mu\nu}-6g_{\mu\nu}-T^{\text{bulk}}_{\mu\nu}=0\,,\label{eq:Einsteineqs}\\
\mathcal{E}_\phi & = & \Box\phi-e^{2\phi}(\partial\chi)^2=0\,,\label{eq:phieom}\\
\mathcal{E}_\chi & = & \Box\chi+2\partial_\mu\phi\partial^\mu\chi=0\,,\label{eq:chieom}
\end{eqnarray}
where
\begin{equation}
T^{\text{bulk}}_{\mu\nu} =
\frac{1}{2}\partial_\mu\phi\partial_\nu\phi+\frac{1}{2}e^{2\phi}\partial_\mu\chi\partial_\nu\chi
-\frac{1}{4}g_{\mu\nu}\left((\partial\phi)^2+\frac{}{}e^{2\phi}(\partial\chi)^2\right)\,.
\end{equation}

The solution expressed as an asymptotic series in radial gauge, i.e. as a Fefferman--Graham (FG) expansion \cite{FeffermanGraham,Graham:1999jg}, reads\footnote{We will denote here and further below by $a_{(n,m)}$ the coefficient at order $r^n(\log r)^m$ of the field $r^\Delta a$ where $r^{-\Delta}$ is the leading term in the expansion of $a$ with the exception of the $a_{(n,0)}$ which we will simply denote as $a_{(n)}$.}
\begin{eqnarray}
g_{\mu\nu}dx^\mu dx^\nu &=& \frac{dr^2}{r^2}+h_{ab}dx^adx^b\,, \label{eq: sol metric gauge} \\
h_{ab} &=& \frac{1}{r^2}\left[h_{(0)ab}+r^2h_{(2)ab}+r^4\log r h_{(4,1)ab}+r^4h_{(4)ab}+\mathcal{O}(r^6\log r)\right]\,, \label{eq: sol metric}\\
\phi &=& \phi_{(0)} + r^2\phi_{(2)}+r^4\log r\phi_{(4,1)} + r^4\phi_{(4)}+\mathcal{O}(r^6\log r)\,, \label{eq: sol phi}\\
\chi &=& \chi_{(0)} + r^2\chi_{(2)}+r^4\log r\chi_{(4,1)}+ r^4\chi_{(4)}+\mathcal{O}(r^6\log r)\label{eq: sol chi}\,,
\end{eqnarray}
where the coefficients are given by
\begin{eqnarray}
h_{(2)ab} &=& -\frac{1}{2}\left(R_{(0)ab}-\frac{1}{2}\partial_a\phi_{(0)}\partial_b\phi_{(0)}-\frac{1}{2}e^{2\phi_{(0)}}\partial_a\chi_{(0)}\partial_b\chi_{(0)}\right) \nonumber\\
&&
+\frac{1}{12}h_{(0)ab}\left(R_{(0)}-\frac{1}{2}(\partial\phi_{(0)})^2-\frac{1}{2}e^{2\phi_{(0)}}(\partial\chi_{(0)})^2\right)\,,\label{eq:h2ab}\\
\phi_{(2)} &=& \frac{1}{4}\left(\square^{(0)}\phi_{(0)}-e^{2\phi_{(0)}}\left(\partial\chi_{(0)}\right)^2\right) \,,\\
\chi_{(2)} &=& \frac{1}{4}\left(\square^{(0)}\chi_{(0)}+2\partial_a\phi_{(0)}\partial^a\chi_{(0)}\right)\label{eq:chi(2)}\,,
\end{eqnarray}
at second order and by
\begin{eqnarray}
h_{(4,1)ab} &=& h_{(2)ac}h^c_{(2)b} + \frac{1}{4}\nabla^{(0)c}\left(\nabla^{(0)}_{a}h_{(2)bc}+\nabla^{(0)}_{b}h_{(2)ac}-\nabla^{(0)}_{c}h_{(2)ab}\right) -\frac{1}{4}\nabla^{(0)}_a\nabla^{(0)}_b h^c_{(2)c} \nonumber\\
&& -\frac{1}{2}\partial_{(a}\phi_{(0)}\nabla^{(0)}_{b)}\phi_{(2)} -
\frac{1}{2}e^{2\phi_{(0)}}\partial_{(a}\chi_{(0)}\nabla^{(0)}_{b)}\chi_{(2)}
- \frac{1}{2}e^{2\phi_{(0)}}
\phi_{(2)}\partial_a\chi_{(0)}\partial_b\chi_{(0)}  \nonumber\\
&& -h_{(0)ab}\left( \frac{1}{4}h_{(2)}^{cd}h_{(2)cd}+\frac{1}{2}\phi_{(2)}^2+\frac{1}{2}e^{2\phi_{(0)}}\chi_{(2)}^2\right) \,,\\
\phi_{(4,1)} &=& -\frac{1}{4}\left[\Box^{(0)}\phi_{(2)}+2\phi_{(2)}h^a_{(2)a} - 4e^{2\phi_{(0)}}\chi^2_{(2)}+\frac{1}{2}\partial^a\phi_{(0)}\nabla^{(0)}_ah^{b}_{(2)b}-h^{ab}_{(2)}\nabla^{(0)}_a\partial_b\phi_{(0)} \right.\nonumber\\
&& \left. -\partial^a\phi_{(0)}\nabla^{(0)b}h_{(2)ab}+e^{2\phi_{(0)}}\partial_a\chi_{(0)}\left(\partial_b\chi_{(0)}h^{ab}_{(2)}-2\phi_{(2)}\partial^a\chi_{(0)}-2\nabla^{(0)a}\chi_{(2)}\right)\right]\,,\nonumber\\
&&\\
\chi_{(4,1)} &=& -\frac{1}{4}\left[ 8\chi_{(2)}\phi_{(2)} +
2\chi_{(2)}h^a_{(2)a} + \Box^{(0)}\chi_{(2)} -
h^{ab}_{(2)}\nabla^{(0)}_a\partial_b\chi_{(0)} +
2\nabla^{(0)}_a\chi_{(2)}\partial^a\phi_{(0)} \right.\nonumber\\
&& \left. +\partial^a\chi_{(0)}\left(\frac{1}{2}
\nabla^{(0)}_ah^b_{(2)b}-\nabla^{(0)b}h_{(2)ab}  -
2\partial^b\phi_{(0)}h_{(2)ab}+
2\nabla^{(0)}_a\phi_{(2)}\right)\right]\,,
\end{eqnarray}
at order $r^4\log r$. We note that $h_{(4,1)ab}$ is traceless. Indices of the expansion coefficients are raised and lowered with the AdS boundary metric $h_{(0)ab}$. At order $r^4$ we have that $h_{(4)ab}$ is constrained by
\begin{eqnarray}
h_{(4)a}^a & = & \frac{1}{4}h_{(2)ab}h_{(2)}^{ab}-\frac{1}{2}\phi_{(2)}^2-\frac{1}{2}e^{2\phi_{(0)}}\chi_{(2)}^2\,,\label{eq:traceh4}\\
\nabla^{(0)b}h_{(4)ab} & = & -e^{2\phi_{(0)}}\chi_{(2)}^2\partial_a\phi_{(0)}+\phi_{(4)}\partial_a\phi_{(0)}+e^{2\phi_{(0)}}\chi_{(4)}\partial_a\chi_{(0)}+e^{2\phi_{(0)}}\phi_{(2)}\chi_{(2)}\partial_a\chi_{(0)}\nonumber\\
&&-\frac{1}{2}\phi_{(2)}\nabla^{(0)}_a\phi_{(2)}-\frac{1}{2}e^{2\phi_{(0)}}\chi_{(2)}\nabla^{(0)}_a\chi_{(2)}-\frac{1}{4}h_{(2)}^{bc}\nabla^{(0)}_ah_{(2)bc}\nonumber\\
&&-\frac{1}{4}h_{(2)ac}\nabla^{(0)c}h_{(2)b}^b+\frac{1}{2}h_{(2)}^{bc}\nabla^{(0)}_bh_{(2)ac}+\frac{1}{2}h_{(2)a}^c\nabla^{(0)b}h_{(2)bc}\,.\label{eq:divh4}
\end{eqnarray}
Following \cite{deHaro:2000xn} we will write $h_{(4)ab}$ as
\begin{equation}
h_{(4)ab} = X_{ab} + \frac{1}{2}t_{ab}\,,\label{eq: h4ab}
\end{equation}
where $t_{ab}$ is the boundary energy-momentum tensor whose trace
and divergence will be given below together with the explicit form
of $X_{ab}$. In the expansion for the scalars we have that $\phi_{(4)}$ and $\chi_{(4)}$ are fully arbitrary functions of the boundary coordinates.

A counterterm action that kills all divergences of the on-shell action $S_{\text{bulk}}+S_{\text{GH}}$ is given by
\begin{eqnarray}
S_{\text{ct}} & = &
\frac{1}{\kappa_5^2}\int_{\partial\mathcal{M}}d^4x\sqrt{-h}\left(-3-\frac{1}{4}Q+\mathcal{A}\left(\lambda+\log r\right)\right)\,,\label{eq:Sct1}
\end{eqnarray}
where $\lambda$ is some scheme dependent parameter (minimal subtraction corresponds to $\lambda=0$) and where
\begin{eqnarray}
\hspace{-1cm}Q\!\! &=&\!\! h^{ab}Q_{ab}\,, \qquad Q_{ab}=R_{(h)ab} - \frac{1}{2}\partial_a\phi\partial_b\phi-\frac{1}{2}e^{2\phi}\partial_a\chi\partial_b\chi\,,\\
\hspace{-1cm}\mathcal{A}\!\! &=&\!\!
\frac{1}{8}\left(Q^{ab}Q_{ab}-\frac{1}{3}Q^2+\frac{1}{2}\left(\square^{(h)}\phi-e^{2\phi}(\partial\chi)^2\right)^2+\frac{1}{2}e^{2\phi}\left(\square^{(h)}\chi+2\partial_a\phi\partial^a\chi\right)^2\right)\,.
\end{eqnarray}
This expression for the conformal anomaly $\mathcal{A}$ differs slightly (by one term) from the expression given in (appendix B of) \cite{Papadimitriou:2011qb}.

\subsection{One-point functions}

We write the total variation of $S_{\text{ren}}$ as
\begin{eqnarray}
\delta S_{\text{ren}} &=&
\frac{1}{2\kappa_5^2}\int_{\mathcal{M}}d^5x\sqrt{-g}\left(
\mathcal{E}_{\mu\nu}\delta g^{\mu\nu} + \mathcal{E}_{\phi}\delta\phi
+ \mathcal{E}_{\chi}\delta\chi \right) \nonumber\\ && -
\frac{1}{2\kappa_5^2} \int_{\partial\mathcal{M}}d^4x\sqrt{-h} \left(
T_{ab}\delta h^{ab} + 2T_{\phi}\delta\phi +
2T_{\chi}\delta\chi \right)\,,
\end{eqnarray}
where $\mathcal{E}_{\mu\nu},\mathcal{E}_{\phi},\mathcal{E}_{\chi}$
are the equations of motion \eqref{eq:Einsteineqs} to \eqref{eq:chieom} and where
\begin{eqnarray}
T_{ab} &=& (K-3)h_{ab} - K_{ab} +
\frac{1}{2}Q_{ab}- \frac{1}{4}h_{ab}Q + \left(\lambda +\log
r\right)T^{(A)}_{ab} \,,\\
T_{\phi} &=& \frac{1}{2}n^\mu\partial_\mu\phi +
\frac{1}{4}\left(\square^{(h)}\phi - e^{2\phi}(\partial\chi)^2\right) +
\left(\lambda +\log
r\right)T^{(A)}_\phi \,,\\
T_{\chi} &=& \frac{1}{2}e^{2\phi}n^\mu\partial_\mu\chi
+\frac{1}{4}e^{2\phi}\left( \square^{(h)}\chi
+2\partial_a\chi\partial^a\phi\right) + \left(\lambda +\log
r\right)T^{(A)}_\chi \,,
\end{eqnarray}
in which we defined
\begin{equation}
T^{(A)}_{ab} = - \frac{2\kappa_5^2}{\sqrt{-h}} \frac{\delta A}{\delta
h^{ab}}\,, \qquad T^{(A)}_{\phi} = - \frac{\kappa_5^2}{\sqrt{-h}}
\frac{\delta A}{\delta \phi}\,,\qquad T^{(A)}_{\chi} = -
\frac{\kappa_5^2}{\sqrt{-h}} \frac{\delta A}{\delta \chi}\,,
\end{equation}
with
\begin{equation}
A=\frac{1}{\kappa_5^2}\int_{\partial\mathcal{M}}d^4x\sqrt{-h}\mathcal{A}\,.
\end{equation}

Using that from the expansions it follows that
$\sqrt{-h}=r^{-4}\sqrt{-h_{(0)}}+\mathcal{O}(r^{-2})$, $\delta
h^{ab} = r^2\delta h_{(0)}^{ab} + \mathcal{O}(r^4)$, $\delta \phi =
\delta \phi_{(0)} + \mathcal{O}(r^2)$ and $\delta \chi = \delta
\chi_{(0)} + \mathcal{O}(r^2)$
we obtain the following one-point functions (we take the cut-off boundary at $r=\epsilon$)
\begin{eqnarray}
\langle T_{(0)ab} \rangle &=& - \frac{2\kappa_5^2}{\sqrt{-h_{(0)}}} \frac{\delta S_{\text{ren}}^{\text{on-shell}}}{\delta h_{(0)}^{ab}} = \lim_{\epsilon \rightarrow 0} \epsilon^{-2} T_{ab} = 2h_{(4)ab} - 2X_{ab}=t_{ab}\,,\label{eq:bdrystresstensor}\\
\langle \mathcal{O}_\phi \rangle &=& - \frac{\kappa_5^2}{\sqrt{-h_{(0)}}} \frac{\delta S_{\text{ren}}^{\text{on-shell}}}{\delta \phi_{(0)}} = \lim_{\epsilon \rightarrow 0} \epsilon^{-4} T_\phi = \nonumber\\
&&-2\phi_{(4)}-\frac{1}{2}\phi_{(2)}h^a_{(2)a} +e^{2\phi_{(0)}}\chi^2_{(2)} -\frac{1}{2} \left(3 - 4\lambda\right)\phi_{(4,1)}\,,\label{eq:vevOdualtophi0}\\
\langle \mathcal{O}_\chi \rangle &=& - \frac{\kappa_5^2}{\sqrt{-h_{(0)}}}
\frac{\delta S_{\text{ren}}^{\text{on-shell}}}{\delta \chi_{(0)}} = \lim_{\epsilon \rightarrow 0}
\epsilon^{-4} T_\chi =\nonumber\\
&& - 2e^{2\phi_{(0)}}\chi_{(4)} -\frac{1}{2}e^{2\phi_{(0)}}\left(\chi_{(2)} h^a_{(2)a} +
4\chi_{(2)}\phi_{(2)} + (3-4\lambda)\chi_{(4,1)}\right) \label{eq:vevOdualtophi0} \,,\qquad
\end{eqnarray}
where
\begin{equation}
X_{ab} = \frac{1}{2}h_{(2)ac}h^c_{(2)b}-\frac{1}{4}h^c_{(2)c}h_{(2)ab} -
\frac{1}{4}h_{(0)ab}\mathcal{A}_{(0)} - \frac{1}{4}\left(3-4\lambda\right)h_{(4,1)ab} \,,\label{eq: Xab}
\end{equation}
with
\begin{equation}
\mathcal{A}_{(0)}=\lim_{\epsilon \rightarrow 0}
\epsilon^{-4}\mathcal{A}
=\frac{1}{2}\left(h_{(2)}^{ab}h_{(2)ab}-(h_{(2)a}^a)^2\right)+\phi_{(2)}^2+e^{2\phi_{(0)}}\chi_{(2)}^2\,.
\end{equation}

The contribution to the one-point functions from the $r^4\log r$ terms in the FG expansions can all be removed by choosing $\lambda = \frac{3}{4}$. The boundary energy-momentum tensor is identified with $t_{ab}$ in \eqref{eq: h4ab}. For any choice of $\lambda$ we compute
its trace and divergence (by using equations \eqref{eq:traceh4} and \eqref{eq:divh4}) and we find
\begin{eqnarray}
t^a_{\;\;a} &=& \mathcal{A}_{(0)}\,,\label{eq:tracet}\\
\nabla^a_{(0)}t_{ab} &=& -\langle \mathcal{O}_\phi \rangle\partial_b\phi_{(0)} -\langle \mathcal{O}_\chi \rangle\partial_b\chi_{(0)} \,.\label{eq:divt}
\end{eqnarray}

\subsection{Manifest $SL(2,\mathbb{R})$ invariance of the counterterm action}

To make the $SL(2,\mathbb{R})$ invariance of the counterterm action $S_{\text{ct}}$ manifest define the matrix of Noether currents (transforming in the adjoint of $SL(2,\mathbb{R})$)
\begin{equation}
\mathcal{J}_\mu=\left(\partial_\mu\mathcal{M}\right)\mathcal{M}^{-1}=\left(\begin{array}{cc}
-J_{(1)\mu} & J_{(3)\mu}\\
J_{(2)\mu} & J_{(1)\mu}
\end{array}\right)\,,
\end{equation}
where $\mathcal{M}$ is given by
\begin{equation}
\mathcal{M}=e^\phi\left(\begin{array}{cc}
\chi^2+e^{-2\phi} & \chi\\
\chi & 1
\end{array}\right)\,.
\end{equation}
We have the three on-shell conserved $SL(2,\mathbb{R})$ Noether currents
\begin{eqnarray}
J_{(1)\mu} & = & \partial_\mu\phi-\chi e^{2\phi}\partial_\mu\chi\,,\\
J_{(2)\mu} & = & e^{2\phi}\partial_\mu\chi\,,\\
J_{(3)\mu} & = & 2\chi\partial_\mu\phi-\chi^2 e^{2\phi}\partial_\mu\chi+\partial_\mu\chi\,.
\end{eqnarray}
The matrix $\mathcal{J}_\mu$ of Noether currents satisfies the properties
\begin{eqnarray}
\nabla^{(h)}_\mu\mathcal{J}_\nu-\nabla^{(h)}_\nu\mathcal{J}_\mu & = & \mathcal{J}_\mu\mathcal{J}_\nu-\mathcal{J}_\nu\mathcal{J}_\mu\,,\\
\mathcal{J}_\mu\mathcal{J}_\nu+\mathcal{J}_\nu\mathcal{J}_\mu & = & \text{Tr}\left(\mathcal{J}_\mu\mathcal{J}_\nu\right)\mathbbm{1}\,.
\end{eqnarray}
The counterterm action \eqref{eq:Sct1} can be rewritten as
\begin{eqnarray}
S_{\text{ct}} & = & \frac{1}{\kappa_5^2}\int_{\partial\mathcal{M}}\!\!\!d^4x\sqrt{-h}\left[-3-\frac{1}{4}Q+\frac{1}{8}\log r\left(Q^{ab}Q_{ab}-\frac{1}{3}Q^2\right.\right.\nonumber\\
&&\left.\left.\hspace{3cm}+\frac{1}{4}\text{Tr}\left(\nabla^{(h)}_a\mathcal{J}^a\nabla^{(h)}_b\mathcal{J}^b\right)\right)\right]\,,\label{eq:Sct}
\end{eqnarray}
where
\begin{equation}
Q_{ab}=R_{(h)ab}-\frac{1}{4}\text{Tr}\left(\mathcal{J}_a\mathcal{J}_b\right)\,,\qquad Q=h^{ab}Q_{ab}\,,
\end{equation}
making manifest its $SL(2,\mathbb{R})$ invariance.

\section{Fefferman--Graham expansions for asymptotically locally $z=2$ Lifshitz space-times}\label{sec:SSreduction}

A pure $z=2$ Lifshitz space-time can be obtained by writing a pure $z=0$ Schr\"odinger space-time in the form of a Kaluza--Klein Ansatz. In order to support the geometry of a $z=0$ Schr\"odinger space-time we need an axionic scalar field. The massive vector field supporting the Lifshitz geometry \cite{Taylor:2008tg} can be obtained by Scherk--Schwarz reduction in which the axion shift symmetry is gauged by the Kaluza--Klein vector\footnote{The 2-form and 3-form matter supporting the Lifshitz geometry that was introduced in \cite{Kachru:2008yh} can be obtained by first dualizing the axion in 5-dimensions to a 3-form potential and then performing an ordinary Kaluza--Klein reduction.}. Hence we can obtain Lagrangians supporting $z=2$ Lifshitz space-times by Scherk--Schwarz reduction of Lagrangians supporting $z=0$ Schr\"odinger space-times \cite{Balasubramanian:2010uk,Donos:2010tu,Cassani:2011sv,Costa:2010cn,Chemissany:2011mb}. We are now in a position to use these observations to peform holographic renormalization for this class of Lagrangians supporting $z=2$ Lifshitz space-times by Scherk--Schwarz reduction using the results of the previous section. 

\subsection{Scherk--Schwarz circle reduction}

We will from now on distinguish between five and four dimensional objects by
putting a hat on all 5-dimensional quantities of the previous section. We split the 5-dimensional coordinates as $x^{\hat\mu}=(x^\mu,u)$.
Consider the following reduction Ansatz
\begin{eqnarray}
d\hat{s}^2 &=& \hat g_{\hat\mu\hat\nu}dx^{\hat\mu}dx^{\hat\nu}=\frac{dr^2}{r^2}+\hat h_{\hat a\hat b}dx^{\hat a}dx^{\hat b}=g_{\mu\nu}dx^\mu dx^\nu + e^{2\Phi}(du+A_\mu dx^\mu)^2\nonumber\\
&=&\frac{dr^2}{r^2}+h_{ab}dx^adx^b+ e^{2\Phi}(du+A_adx^a)^2\,,\label{eq:metricdecomposition5Dto4D}\\
\hat{\chi} &=& \chi + ku \,,\label{eq: decomposition of chi}\\
\hat{\phi} &=& \phi \label{eq: decomposition of phi}\,,
\end{eqnarray}
where the four dimensional unhatted fields are all independent of the fifth coordinate $u$ which is periodically identified as $u\sim u+2\pi L$. The reduced theory expressed in terms of the 4-dimensional metric $g_{\mu\nu}$ will not be in Einstein frame. The frame in \eqref{eq:metricdecomposition5Dto4D} is such that we preserve the 5-dimensional radial gauge \eqref{eq: sol metric gauge} in four dimensions. We will perform the holographic renormalization of the reduced 4-dimensional theory in this frame. This construction is very reminiscent of the methods used in \cite{Kanitscheider:2008kd,Gouteraux:2011qh} in the case of dimensional reduction from asymptotically locally AdS space-times to space-times that are (in Einstein frame) asymptotically conformally locally AdS. 

For the dimensional reduction of \eqref{eq: Bulk lagrangian} to \eqref{eq:GHaction} the following relations are useful
\begin{eqnarray}
\sqrt{-\hat g} & = & e^{\Phi}\sqrt{-g}\,,\\
\hat n^\mu & = & n^\mu\,,\\
\sqrt{-\hat h} & = &  e^\Phi\sqrt{-h}\,,\\
\hat K & = & K+n^\mu\partial_\mu\Phi\,,\\
\hat R & = & R-2\square\Phi-2\left(\partial\Phi\right)^2-\frac{1}{4}e^{2\Phi}F^2\,.
\end{eqnarray}
Using these relations we find
\begin{eqnarray}
S & = & \frac{1}{2\kappa_5^2}\int d^5x\sqrt{-\hat g}\left(\hat R+12-\frac{1}{2}(\partial\hat\phi)^2-\frac{1}{2}e^{2\hat\phi}\left(\partial\hat\chi\right)^2\right)\nonumber\\
&&+\frac{1}{\kappa_5^2}\int d^4x\sqrt{-\hat h}\hat K+S_{\text{ct}}\nonumber\\
& = & \frac{2\pi L}{2\kappa_5^2}\int d^4x\sqrt{-g}\left(e^\Phi R-\frac{1}{4}e^{3\Phi}F^2-\frac{1}{2}e^\Phi\left(\partial\phi\right)^2-\frac{1}{2}e^{\Phi+2\phi}\left(\mathcal{D}\chi\right)^2-e^{2\Phi}V\right)\nonumber\\
&&+\frac{2\pi L}{\kappa_5^2}\int d^3x\sqrt{-h}e^\Phi K+S_{\text{ct}}\,,\label{eq:4Daction}
\end{eqnarray}
where
\begin{eqnarray}
\mathcal{D}_\mu\chi & = & \partial_\mu\chi-kA_\mu\equiv -kB_\mu\,,\label{eq:B}\\
F_{\mu\nu} & = & \partial_\mu A_\nu-\partial_\nu A_\mu=\partial_\mu B_\nu-\partial_\nu B_\mu\,,\\
V & = & \frac{k^2}{2}e^{-3\Phi+2\phi}-12e^{-\Phi}\,,
\end{eqnarray}
in which $B_\mu$ is the massive vector field (that only exists for $k\neq 0$) and where $S_{\text{ct}}$ is a counterterm action. From now on we will take $k\neq 0$ and replace $\mathcal{D}_\mu\chi$ by $-kB_\mu$ giving
\begin{eqnarray}
S & = & \frac{2\pi L}{2\kappa_5^2}\int d^4x\sqrt{-g}\left(e^\Phi R-\frac{1}{4}e^{3\Phi}F^2-\frac{1}{2}e^\Phi\left(\partial\phi\right)^2-\frac{k^2}{2}e^{\Phi+2\phi}B^2-e^{2\Phi}V\right)\nonumber\\
&&+\frac{2\pi L}{\kappa_5^2}\int d^3x\sqrt{-h}e^\Phi K+S_{\text{ct}}\,,\label{eq:4DactionB}
\end{eqnarray}
where
\begin{equation}\label{eq:V}
V=\frac{k^2}{2}e^{-3\Phi+2\phi}-12e^{-\Phi}\,.
\end{equation}
The 4-dimensional equations of motion associated with the action \eqref{eq:4DactionB} are
\begin{eqnarray}
R_{\mu\nu} & = & \nabla_\mu\partial_\nu\Phi+\frac{1}{2}g_{\mu\nu}\square\Phi+\partial_\mu\Phi\partial_\nu\Phi+\frac{1}{2}g_{\mu\nu} \left(\partial\Phi\right)^2+\frac{1}{2}g_{\mu\nu}e^{\Phi}V\nonumber\\
&&+\frac{1}{2}\partial_\mu\phi\partial_\nu\phi+\frac{k^2}{2}e^{2\phi}B_\mu B_\nu+\frac{1}{2}e^{2\Phi}\left(F_{\mu\rho}F_{\nu}{}^\rho-\frac{1}{4}g_{\mu\nu}F^2\right)\,,\label{eq:4DEinsteineqs}\\
0 & = & \square\Phi+\left(\partial\Phi\right)^2-\frac{1}{4}e^{2\Phi}F^2+\frac{k^2}{2}e^{-2\Phi+2\phi}-4\,,\\
0 & = & \nabla_\mu\left(e^\Phi \partial^\mu\phi\right)-k^2e^{2\phi+\Phi}B^2-k^2e^{-\Phi+2\phi}\,,\\
0 & = & \nabla_\mu\left(e^{3\Phi}F^{\mu\nu}\right)-k^2e^{2\phi+\Phi}B^\nu\,.\label{eq:Aeom}
\end{eqnarray}

\subsection{The $z=2$ Lifshitz space-time}

The equations \eqref{eq:4DEinsteineqs} to \eqref{eq:Aeom} admit the pure $z=2$ Lifshitz space-time as a solution,
\begin{eqnarray}
ds^2 & = & \frac{dr^2}{r^2}-e^{-2\Phi_{(0)}}\frac{dt^2}{r^4}+\frac{1}{r^2}\left(dx^2+dy^2\right)\,,\label{eq:Lifshitzmetric}\\
B & = & -e^{-2\Phi_{(0)}}\frac{dt}{r^2}\,,\\
\Phi & = & \Phi_{(0)}=\phi_{(0)}+\log\frac{k}{2}\,,\label{eq:PhipureLifshitz}\\
\phi & = & \phi_{(0)}=\text{cst}\,.
\end{eqnarray}
From a 5-dimensional perspective this solution is a $z=0$ Schr\"odinger space-time and reads
\begin{eqnarray}
d\hat s^2 & = & \frac{dr^2}{r^2}+\frac{1}{r^2}\left(-2dtdu+dx^2+dy^2\right)+\frac{k^2}{4}e^{2\hat\phi_{(0)}}du^2\,,\label{eq:Schz=0}\\
\hat\phi & = & \hat\phi_{(0)}=\phi_{(0)}=\text{cst}\,,\\
\hat\chi & = & ku+\text{cst}\,.
\end{eqnarray}

Before studying more general solutions that asymptote to \eqref{eq:Lifshitzmetric} in a certain sense we will first study the pure Lifshitz space-time from a 5-dimensional point of view using an arbitrary Fefferman--Graham coordinate system where we only keep manifest the $u$ coordinate for the purpose of performing the Scherk--Schwarz reduction. In the language of the previous section this means that we should take
\begin{eqnarray}
\hat\phi_{(0)} & = & \text{cst}\,,\label{eq:Lifshitz1}\\
\hat\phi_{(4)} & = & 0\,,\\
\hat\chi_{(0)} & = & ku+\text{cst}\,,\\
\hat\chi_{(4)} & = & 0\,,\\
\hat h_{(0)\hat a\hat b} & = & \text{conformally flat and admits a hypersurface} \nonumber\\
&& \text{orthogonal null Killing vector $\partial_u$}\,,\\
\hat t_{\hat a\hat b} & = & 0\,.\label{eq:Lifshitz6}
\end{eqnarray}
The properties of $\hat h_{(0)\hat a\hat b}$ can be easily understood. From the reduction Ansatz \eqref{eq:metricdecomposition5Dto4D} we learn that
\begin{equation}
e^{2\Phi} = \hat h_{uu}\,.\label{eq:Phi}
\end{equation}
In order that $\Phi$ is a constant it is necessary that
\begin{equation}\label{eq:null}
\hat h_{(0)uu}=0\,.
\end{equation}
Since in order to do the reduction we need that $\partial_u$ is a Killing vector of the 5-dimensional metric and because $u$ is a boundary coordinate we find that $\partial_u$ is a null Killing vector of the boundary metric. Further we also need that the boundary value of $\Phi$ is fixed by \eqref{eq:PhipureLifshitz}. This requires that
\begin{equation}
\hat h_{(2)uu}=\frac{k^2}{4}e^{2\hat\phi_{(0)}}\,.
\end{equation}
This in turn is only possible provided we impose (as follows from \eqref{eq:h2ab})
\begin{equation}\label{eq:HSO}
\hat R_{(0)uu}=0\,.
\end{equation}
Using that $\partial_u$ is a null Killing vector and thus tangent to a null geodesic congruence it will be shown below, with the help of the Raychaudhuri equation, that provided \eqref{eq:HSO} holds, the null Killing vector $\partial_u$ is hypersurface orthogonal. Finally since the metric \eqref{eq:Schz=0} is asymptotically AdS it follows that the AdS boundary metric is conformally flat. This explains the condition imposed on $\hat h_{(0)\hat a\hat b}$ which together with \eqref{eq:Lifshitz6} imply that $\Phi$ is a constant satisfying \eqref{eq:PhipureLifshitz}. Further these conditions for $\hat h_{(0)\hat a\hat b}$ combined with \eqref{eq:Lifshitz6} are necessary and sufficient in order that the 5-dimensional metric is of the form 
\begin{equation}
d\hat s^2=ds^2_{\text{AdS}}+\frac{k^2}{4}e^{2\hat\phi_{(0)}}du^2\,,
\end{equation}
where $ds^2_{\text{AdS}}$ is the metric of a pure 5-dimensional AdS space-time. This completes the 5-dimensional uplift of the pure Lifshitz metric in Fefferman--Graham coordinates.

\subsection{Boundary parametrizations}

In subsection \ref{subsec:AlLif} we will generalize the results of the previous subsection to the case of AlLif space-times defined from a 5-dimensional point of view. In order to prepare for that we will now discuss one of the conditions that goes into the definition of AlLif space-times defined from a 5-dimensional point of view as this will provide guidance for how to proceed. For the pure Lifshitz solution the value of $\Phi-\phi$ was not a free parameter but equal to $\log\frac{k}{2}$. Since it is not a free parameter we will require that this is still true for AlLif space-times. This is enough to deduce that once again $\hat h_{(0)\hat a\hat b}$ must admit a hypersurface orthogonal null Killing vector. To see this one just observes that we again need equations \eqref{eq:null} and \eqref{eq:HSO} and of course that $\partial_u$ is a Killing vector of the complete 5-dimensional metric. We do not require that $\hat h_{(0)\hat a\hat b}$ is conformally flat so that an AlLif space-time uplifts to an asymptotically locally AdS space-time as defined in \cite{Papadimitriou:2005ii}.

Since we will always require that $\hat h_{(0)\hat a\hat b}$ admits a hypersurface orthogonal null Killing vector it will be useful to consider the following double null split of the boundary metric
\begin{equation}
\hat h_{(0)\hat a\hat b}=-\hat N_{(0)\hat a}\hat H_{(0)\hat b}-\hat N_{(0)\hat b}\hat H_{(0)\hat a}+\hat\Pi_{(0)\hat a\hat b}\,,
\end{equation}
where, say, $\hat H^{\hat a}_{(0)}$ is identified with the null Killing direction $\partial_u$ and where $\hat N_{(0)}^{\hat a}$ is a second null vector satisfying
\begin{equation}\label{eq:nullbein}
\hat N_{(0)\hat a}\hat H_{(0)}^{\hat a}=-1\,.
\end{equation}
Further we impose that $\hat\Pi_{(0)\hat a\hat b}$ is a projector onto a 2-dimensional Euclidean subspace orthogonal to both $\hat N_{(0)}^{\hat a}$ and $\hat H_{(0)}^{\hat a}$. The indices on $\hat H_{(0)\hat a}$, $\hat N_{(0)\hat a}$ and $\hat\Pi_{(0)\hat a\hat b}$ are raised and lowered using
$\hat h_{(0)\hat a\hat b}$.

We will now use the above introduced notation to show that a null Killing vector field $\hat H_{(0)}=\partial_u$ satisfying \eqref{eq:HSO} is indeed hypersurface orthogonal. The vector $\hat H^{\hat a}_{(0)}$ being a null Killing vector is tangent to a null geodesic congruence. Define
\begin{equation}
\hat B_{(0)\hat a\hat b}=\hat\nabla^{(0)}_{\hat a}\hat H_{(0)\hat b}\,,
\end{equation}
as well as 
\begin{equation}
\hat S_{(0)\hat a\hat b}=\hat\Pi_{(0)\hat a}{}^{\hat c}\hat\Pi_{(0)\hat b}{}^{\hat d}\hat B_{(0)\hat c\hat d}\,,
\end{equation}
which is the projected version of $\hat B_{(0)\hat a\hat b}$ with the projection onto the co-dimension two subspace orthogonal to both $\hat N_{(0)}^{\hat a}$ and $\hat H_{(0)}^{\hat b}$. This space is not uniquely defined as $\hat N^{\hat a}_{(0)}$, being only constrained by \eqref{eq:nullbein}, is not uniquely defined. Anyway, the results will not depend on the specific choice for $\hat N^{\hat a}_{(0)}$. Because $\hat H^{\hat a}_{(0)}$ is Killing, the shear and expansion of the null geodesic congruence are zero and the Raychaudhuri equation reads
\begin{equation}
\hat H^{\hat a}_{(0)}\hat H^{\hat b}_{(0)}\hat R_{(0)\hat a\hat b}=\hat\omega_{(0)\hat a\hat b}\hat\omega_{(0)}^{\hat a\hat b}\,,
\end{equation}
where 
\begin{equation}
\hat\omega_{(0)\hat a\hat b}=\hat S_{(0)[\hat a\hat b]}\,.
\end{equation}
Hence, whenever \eqref{eq:HSO} holds we have for a null Killing vector that
\begin{equation}
\hat\omega_{(0)\hat a\hat b}\hat\omega^{\hat a\hat b}_{(0)}=0\quad\rightarrow\quad\hat\omega_{(0)\hat a\hat b}=0\,.
\end{equation}
We now show that this implies that $\hat H^{\hat a}_{(0)}$ is hypersurface orthogonal. We have from the definition of $\hat\omega_{(0)\hat a\hat b}$ and the properties of $\hat H^{\hat a}_{(0)}$ that
\begin{equation}
d\hat H_{(0)}=\hat H_{(0)}\wedge\hat V_{(0)}\,,
\end{equation}
where $\hat V_{(0)\hat a}=2\hat N^{\hat b}_{(0)}\hat\nabla^{(0)}_{[\hat a}\hat H_{(0)\hat b]}$. It follows that for arbitrary $\hat V_{(0)\hat a}$ and hence for any choice of $\hat N^{\hat a}_{(0)}$ that
\begin{equation}
\hat H_{(0)}\wedge d\hat H_{(0)}=0\,,
\end{equation}
which is the Frobenius integrability condition for $\hat H^{\hat a}_{(0)}$ to be hypersurface orthogonal.

Because $\hat H_{(0)}^{\hat a}$ is hypersurface orthogonal we can always (locally) choose coordinates such that
\begin{equation}\label{eq:HSOadaptedcoordinates}
\hat H_{(0)\hat a}=H_{(0)}\partial_{\hat a}t\,.
\end{equation}
Further the only condition imposed on $\hat N_{(0)}^{\hat a}$ is that it satisfies \eqref{eq:nullbein}. Hence we can assume without loss of generality that also $\hat N_{(0)}^{\hat a}$ is hypersurface orthogonal and given by
\begin{equation}
 \hat N_{(0)\hat a}=N_{(0)}\partial_{\hat a}u\,.
\end{equation}
It follows that $N_{(0)}=-1$ and $\hat H_{(0)}^i=0$. Hence we thus have
\begin{equation}\label{eq:bdrymetricparametrization}
\hat h_{(0)\hat a\hat b}dx^{\hat a}dx^{\hat b}=2H_{(0)}dudt+\Pi_{(0)ij}\left(dx^i+H_{(0)}N_{(0)}^idt\right)\left(dx^j+H_{(0)}N_{(0)}^jdt\right)\,,
\end{equation}
where we dropped the hat on $\hat\Pi_{(0)ij}$ and on $\hat N_{(0)}^i$ and where all metric components are arbitrary functions of $t$ and $x^i$ but do not depend on $u$. In appendix \ref{eq:appnullsplit} we provide some explicit formulas for the geometric quantities of interest depending on $\hat h_{(0)\hat a\hat b}$ expressed in the coordinate system \eqref{eq:bdrymetricparametrization}.

\subsection{Dimensional reduction of the Fefferman--Graham expansions}

From the reduction Ansatz \eqref{eq:metricdecomposition5Dto4D} and \eqref{eq: decomposition of chi} together with \eqref{eq:B} it follows that
in radial gauge for $\hat g_{\hat\mu\hat\nu}$ we have the following relation between the 4- and 5-dimensional fields
\begin{eqnarray}
g_{rr} & = & \frac{1}{r^2}\,,\\
g_{ra} & = & 0\,,\\
h_{ab} & = & \hat h_{ab}-\frac{\hat h_{au}\hat h_{bu}}{\hat h_{uu}}\,,\label{eq:hab}\\
B_r & = & -\frac{1}{k}\partial_r\hat\chi\,,\\
B_a & = & \frac{\hat h_{a u}}{\hat h_{uu}}-\frac{1}{k}\partial_a\hat\chi\,,\label{eq:Ba}\\
\Phi & = & \frac{1}{2}\log\hat h_{uu}\,,\label{eq:Phi}
\end{eqnarray}
where the hatted fields satisfy the reduction Ansatz. The double null split of the 5-dimensional boundary metric $\hat h_{(0)\hat a\hat b}$ puts the 3-dimensional metric $h_{ab}$, defined in \eqref{eq:hab}, in ADM form. We will work out the expansions of the 4-dimensional fields assuming only that $\hat h_{(0)\hat a\hat b}$ is parametrized as in \eqref{eq:bdrymetricparametrization}.

Using equations \eqref{eq:hab} to \eqref{eq:Phi} together with \eqref{eq: decomposition of phi} as well as the 5-dimensional expansions \eqref{eq: sol metric} to \eqref{eq: sol chi} and the boundary parametrization \eqref{eq:bdrymetricparametrization}, we obtain the following expansions for the 4-dimensional fields\footnote{We note that the $r$ component of the massive vector field goes to zero as we approach the boundary. This boundary condition is very similar to what has been proposed for asymptotically Schr\"odinger space-times in \cite{Hartong:2012sw}.}
\begin{eqnarray}
h_{tt} & = & \frac{1}{r^4}\left(h_{(0)tt} +r^2\log r h_{(2,1)tt}+
r^2h_{(2)tt}+r^4(\log r)^2 h_{(4,2)tt}\right.\nonumber\\
&&\left.+r^4\log r h_{(4,1)tt}+r^4h_{(4)tt}+\mathcal{O}(r^6(\log r)^3)\right)\,,\label{eq:htt}\\
h_{ti} & = & \frac{1}{r^2}\left(h_{(0)ti}+r^2\log r h_{(2,1)ti}+r^2h_{(2)ti}+\mathcal{O}(r^4(\log r)^2)\right)\,,\label{eq:hit}\\
h_{ij} & = & \frac{1}{r^2}\left(h_{(0)ij}+r^2h_{(2)ij}+r^4\log r h_{(4,1)ij}+r^4h_{(4)ij}+\mathcal{O}(r^6(\log r)^2)\right)\,,\label{eq:hij}\\
B_r & = & r\left(B_{(0)r}+r^2\log rB_{(2,1)r}+r^2B_{(2)r}+\mathcal{O}(r^4\log r)\right)\,,\\
B_t  & = &  \frac{1}{r^2}\left(B_{(0)t} +r^2\log r B_{(2,1)t}+r^2B_{(2)t}+r^4(\log r)^2 B_{(4,2)t}\right.\nonumber\\
&&\left.+r^4\log r B_{(4,1)t}+r^4B_{(4)t}+\mathcal{O}(r^6(\log r)^3)\right)\,,\\
B_i  & = &  B_{(0)i}+r^2\log r B_{(2,1)i}+r^2B_{(2)i}+\mathcal{O}(r^4(\log r)^2)\,,\\
\Phi &  = & \Phi_{(0)} + r^2\log r\Phi_{(2,1)}+ r^2\Phi_{(2)} +r^4(\log r)^2\Phi_{(4,2)}+r^4\log r\Phi_{(4,1)}\nonumber\\
&&+r^4\Phi_{(4)}+\mathcal{O}(r^6(\log r)^3)\,,\\
\phi & = & \phi_{(0)}+r^2\phi_{(2)}+r^4\log r\phi_{(4,1)}+r^4\phi_{(4)}+\mathcal{O}(r^6\log r)\,.\label{eq:phi}
\end{eqnarray}
For the determinant and inverse metric we get the following expansions
\begin{eqnarray}
\sqrt{-h} & = & e^{-\Phi}\sqrt{\hat h}=r^{-4}\sqrt{-h_{(0)tt}}\sqrt{\Pi_{(0)}}\left(1+r^2\log r h_{(2,1)}+r^2h_{(2)}\right.\nonumber\\
&&\left.+r^4(\log r)^2 h_{(4,2)}+r^4\log r h_{(4,1)}+r^4h_{(4)}+\mathcal{O}\left(r^6(\log r)^3\right)\right)\,,\\
h^{tt} & = & \hat h^{tt}=r^4\left(s_{(0)}^{tt}+r^2\log r s_{(2,1)}^{tt}+r^2s_{(2)}^{tt}+\mathcal{O}(r^4(\log r)^2)\right)\,,\label{eq:invhtt}\\
h^{ti} & = & \hat h^{ti}=r^4\left(s_{(0)}^{ti}+r^2\log r s_{(2,1)}^{ti}+r^2s_{(2)}^{ti}+\mathcal{O}(r^4(\log r)^2)\right)\,,\label{eq:inhti}\\
h^{ij} & = & \hat h^{ij}=r^2\left(s_{(0)}^{ij}+r^2s_{(2)}^{ij}+\mathcal{O}(r^4\log r)\right)\,,\label{eq:invhij}
\end{eqnarray}
where
\begin{eqnarray}
h_{(2,1)} & = & \frac{1}{2}\frac{h_{(2,1)tt}}{h_{(0)tt}}\,,\\
h_{(2)} & = & \frac{1}{2}\left(\frac{h_{(2)tt}}{h_{(0)tt}}-\Pi_{(0)}^{kl}\frac{h_{(0)tk}h_{(0)tl}}{h_{(0)tt}}+\Pi_{(0)}^{kl}h_{(2)kl}\right)\,,\\
s_{(0)}^{tt} & = & \frac{1}{h_{(0)tt}}\,,\\
s_{(0)}^{ti} & = & -\Pi_{(0)}^{ij}\frac{h_{(0)tj}}{h_{(0)tt}}\,,\\
s_{(0)}^{ij} & = & \Pi_{(0)}^{ij}\,,
\end{eqnarray}
\begin{eqnarray}
s_{(2,1)}^{tt} & = & -\frac{h_{(2,1)tt}}{(h_{(0)tt})^2}\,,\\
s_{(2,1)}^{ti} & = & \Pi_{(0)}^{ij}\left[-\frac{h_{(2,1)tj}}{h_{(0)tt}}+\frac{h_{(2,1)tt}h_{(0)tj}}{(h_{(0)tt})^2}\right]\,,\\
s_{(2)}^{tt} & = & -\frac{h_{(2)tt}}{(h_{(0)tt})^2}+\Pi_{(0)}^{ij}\frac{h_{(0)ti}h_{(0)tj}}{(h_{(0)tt})^2}\,,\\
s_{(2)}^{ti} & = & \Pi_{(0)}^{ij}\left[-\frac{h_{(2)tj}}{h_{(0)tt}}+\frac{h_{(0)tj}h_{(2)tt}}{(h_{(0)tt})^2}-\Pi_{(0)}^{kl}\frac{h_{(0)tj}h_{(0)tk}h_{(0)tl}}{(h_{(0)tt})^2}\right.\nonumber\\
&&\left.+\Pi_{(0)}^{kl}\frac{h_{(0)tk}h_{(2)jl}}{h_{(0)tt}}\right]\,,\\
s_{(2)}^{ij} & = & \Pi_{(0)}^{ik}\Pi_{(0)}^{jl}\left(-h_{(2)kl}+\frac{h_{(0)tk}h_{(0)tl}}{h_{(0)tt}}\right)\,.
\end{eqnarray}

We next work out the coefficients appearing in the above expansions \eqref{eq:htt} to \eqref{eq:phi} up to second order. From plugging in \eqref{eq: sol metric} to \eqref{eq: sol chi} into \eqref{eq:hab} to \eqref{eq:Phi} we find at leading order the following relations
\begin{eqnarray}
h_{(0)tt} & = & -(B_{(0)t})^2e^{2\Phi_{(0)}}\,,\label{eq:constraint0thorder}\\
\Phi_{(0)} & = & \phi_{(0)}+\log\frac{k}{2}\,.
\end{eqnarray}
The constraint \eqref{eq:constraint0thorder} is a consequence of the fact that the Killing vector $\partial_u$ is null on the boundary of the 5-dimensional asymptotically locally AdS space-time. We will parametrize the constraint equation \eqref{eq:constraint0thorder} by writing
\begin{eqnarray}
B_{(0)t} & = & H_{(0)}e^{-2\Phi_{(0)}}\,,\\
h_{(0)tt} & = & -H_{(0)}^2e^{-2\Phi_{(0)}}\,,
\end{eqnarray}
where $H_{(0)}$ originates from the reduction ($\hat h_{(0)tu}=H_{(0)}$) and we eliminate $\phi_{(0)}$ in favor of $\Phi_{(0)}$. Further since we have $h_{(0)ij}=\Pi_{(0)ij}$ we will use $\Pi_{(0)ij}$ in the expansions.

In order to express the 4-dimensional expansions that are obtained by the reduction procedure in terms of the coefficients appearing in \eqref{eq:htt} to \eqref{eq:phi} we first work out the expansion expressed in terms of the Kaluza--Klein vector field $A_\mu$ and the axion $\chi$. The result of this analysis is given in appendix \ref{app:4DexpansionsAandchi}. In order to re-express the result of appendix \ref{app:4DexpansionsAandchi} in terms of the coefficients of the massive vector field $B_\mu$ we go about as follows. Replace $\partial_i\chi_{(0)}$ using that $h_{(0)ti}=H_{(0)}(N_{(0)i}-\frac{1}{k}\partial_i\chi_{(0)})$. This leads to an explicit dependence on $N_{(0)i}$ which we do not want. However, it turns out that $N_{(0)i}$ only appears as $N_{(0)i}N_{(0)}^i$ and this can be removed using equation \eqref{eq:h2ttapp} which can be written as
\begin{equation}\label{eq:Nsquaredtoh2tt}
H_{(0)}^2N_{(0)i}N_{(0)}^i=h_{(2)tt}+2H_{(0)}A_{(2)t}+2H_{(0)}^2e^{-2\Phi_{(0)}}\Phi_{(2)}\,.
\end{equation}
Once we have removed $\partial_i\chi_{(0)}$ and $N_{(0)i}$ from the expansions of appendix \ref{app:4DexpansionsAandchi} in this way the result can be easily expressed in terms of the coefficients of the massive vector field $B_{\mu}$ using that
\begin{eqnarray}
B_{(0)t} & = & A_{(0)t}\,,\label{eq:B0t}\\
B_{(2,1)t} & = & A_{(2,1)t}\,,\\
B_{(2)t} & = & A_{(2)t}-\frac{1}{k}\partial_t\chi_{(0)}\,,\\
B_{(0)i} & = & A_{(0)i}-\frac{1}{k}\partial_i\chi_{(0)}\,,\\
B_{(2)i} & = & A_{(2)i}-\frac{1}{k}\partial_i\chi_{(2)}\,,\\
B_{(0)r} & = & -\frac{2}{k}\chi_{(2)}\,.\label{eq:B0r}
\end{eqnarray}
We end up with the following expressions
\begin{eqnarray}
h_{(2,1)tt} & = & 2e^{-2\Phi_{(0)}}H_{(0)}^2\Phi_{(2,1)}\,,\\
h_{(2,1)ti} & = & -H_{(0)}B_{(2,1)i}\,,\\
h_{(2)ti} & = & e^{2\Phi_{(0)}}H_{(0)}^{-1}h_{(0)ti}B_{(2)t}+2h_{(0)ti}\Phi_{(2)}-H_{(0)}B_{(2)i}+\frac{1}{2}H_{(0)}\partial_iB_{(0)r}\nonumber\\
&&+\frac{1}{4}\partial_t\Phi_{(0)}\partial_i\Phi_{(0)}+\frac{1}{4}\Pi_{(0)}^{jk}\left(D^{(0)}_i\partial_t\Pi_{(0)jk}-D^{(0)}_j\partial_t\Pi_{(0)ik}\right)+\frac{1}{4}H_{(0)}^{-1}\partial_i\partial_t H_{(0)}\nonumber\\
&&-\frac{1}{4}H_{(0)}^{-2}\partial_t H_{(0)}\partial_iH_{(0)}-\frac{1}{8}H_{(0)}^{-1}\partial_iH_{(0)}\Pi_{(0)}^{jk}\partial_t\Pi_{(0)jk}+\frac{1}{4}D^{(0)j} D^{(0)}_j h_{(0)ti}\nonumber\\
&&-\frac{1}{4}D^{(0)j} D^{(0)}_i h_{(0)tj}+\frac{1}{4}H_{(0)}^{-1}h_{(0)t}{}^jD^{(0)}_j\partial_i H_{(0)}-\frac{1}{4}h_{(0)ti}H_{(0)}^{-1}D^{(0)j}\partial_jH_{(0)}\nonumber\\
&&+\frac{1}{4}H_{(0)}^{-1}\partial_i H_{(0)}D^{(0)j} h_{(0)tj}-\frac{1}{4}H_{(0)}^{-1}\left(\partial^j H_{(0)}\right)D^{(0)}_i h_{(0)tj}\,,\\
h_{(2)ij} & = & -\frac{1}{2}\mathcal{R}_{(0)ij}+\frac{1}{2}H_{(0)}^{-1}D^{(0)}_i\partial_jH_{(0)}-\frac{1}{4}H_{(0)}^{-2}\partial_iH_{(0)}\partial_jH_{(0)}+\frac{1}{4}\partial_i\Phi_{(0)}\partial_j\Phi_{(0)}\nonumber\\
&&+\Pi_{(0)ij}\left(e^{2\Phi_{(0)}}H_{(0)}^{-1}B_{(2)t}+2\Phi_{(2)}-\frac{1}{4}H_{(0)}^{-1}D^{(0)k}\partial_kH_{(0)}\right)\,,\\
B_{(0)r} & = & \frac{1}{2}H_{(0)}^{-1}D^{(0)}_i h_{(0)t}{}^i-\frac{1}{4}H_{(0)}^{-1}\Pi_{(0)}^{ij}\partial_t\Pi_{(0)ij}+H_{(0)}^{-1}h_{(0)ti}\partial^i\Phi_{(0)}\nonumber\\
&&-H_{(0)}^{-1}\partial_t\Phi_{(0)}\,,\\
B_{(2,1)t} & = & -2e^{-2\Phi}H_{(0)}\Phi_{(2,1)}\,,\\
B_{(2)t} & = & \frac{1}{8}e^{-2\Phi_{(0)}}H_{(0)}\left(\mathcal{R}_{(0)}+H_{(0)}^{-1}D^{(0)i}\partial_iH_{(0)}+\frac{1}{2}H_{(0)}^{-2}\partial_iH_{(0)}\partial^iH_{(0)}\right.\nonumber\\
&&\left.-\frac{1}{2}\partial_i\Phi_{(0)}\partial^i\Phi_{(0)}-20\Phi_{(2)}\right)+\frac{1}{4}H_{(0)}^{-1}h_{(2)tt}-\frac{1}{4}H_{(0)}^{-1}h_{(0)ti}h_{(0)t}{}^i\,,\label{eq:B2t}\\
B_{(0)i} & = & 0\,,
\end{eqnarray}
\begin{eqnarray}
B_{(2,1)i} & = & \frac{1}{4}H_{(0)}^{-1}\mathcal{R}_{(0)ij}h_{(0)t}{}^j+\frac{1}{4}H_{(0)}^{-1}D^{(0)j}D^{(0)}_jh_{(0)ti}+\frac{1}{4}H_{(0)}^{-2}\partial_iH_{(0)}D^{(0)}_jh_{(0)t}{}^j\nonumber\\
&&-\frac{1}{4}H_{(0)}^{-2}\partial^jH_{(0)}\left(D^{(0)}_ih_{(0)tj}+D^{(0)}_jh_{(0)ti}\right)-\frac{1}{2}H_{(0)}^{-1}D^{(0)}_jh_{(0)t}{}^j\partial_i\Phi_{(0)}\nonumber\\
&&+\frac{1}{2}H_{(0)}^{-1}\partial^j\Phi_{(0)}\left(D^{(0)}_ih_{(0)tj}+D^{(0)}_jh_{(0)ti}\right)-\frac{5}{4}H_{(0)}^{-1}h_{(0)t}{}^j\partial_i\Phi_{(0)}\partial_j\Phi_{(0)}\nonumber\\
&&-\frac{1}{4}H_{(0)}^{-1}D^{(0)j}\partial_t\Pi_{(0)ij}+\frac{1}{8}H_{(0)}^{-1}\Pi_{(0)}^{jk}D^{(0)}_i\partial_t\Pi_{(0)jk}+\frac{1}{4}H_{(0)}^{-2}\partial^jH_{(0)}\partial_t\Pi_{(0)ij}\nonumber\\
&&-\frac{1}{8}H_{(0)}^{-2}\partial_iH_{(0)}\Pi_{(0)}^{jk}\partial_t\Pi_{(0)jk}+\frac{1}{4}H_{(0)}^{-1}\partial_i\Phi_{(0)}\Pi_{(0)}^{jk}\partial_t\Pi_{(0)jk}\nonumber\\
&&-\frac{1}{2}H_{(0)}^{-1}\partial^j\Phi_{(0)}\partial_t\Pi_{(0)ij}+\frac{5}{4}H_{(0)}^{-1}\partial_t\Phi_{(0)}\partial_i\Phi_{(0)}\,,\\
\Phi_{(2,1)} & = & -\frac{8}{3}\Phi_{(2)}-\frac{8}{3}e^{2\Phi_{(0)}}H_{(0)}^{-1}B_{(2)t}-\frac{4}{3}e^{2\Phi_{(0)}}H_{(0)}^{-2}h_{(2)tt}-\frac{1}{2}D^{(0)i}\partial_i\Phi_{(0)}\nonumber\\
&&-\frac{13}{24}\partial_i\Phi_{(0)}\partial^i\Phi_{(0)}+\frac{4}{3}e^{2\Phi_{(0)}}H_{(0)}^{-2}h_{(0)ti}h_{(0)t}{}^i\nonumber\\
&&+\frac{1}{12}\left(\mathcal{R}_{(0)}+D^{(0)i}\partial_i\log H_{(0)}\right)\,,\\
\phi_{(2)} & = & 2e^{2\Phi_{(0)}}H_{(0)}^{-1}B_{(2)t}+e^{2\Phi_{(0)}}H_{(0)}^{-2}h_{(2)tt}+2\Phi_{(2)}+\frac{1}{4}D^{(0)i}\partial_i\Phi_{(0)}\nonumber\\
&&+\frac{1}{4}H_{(0)}^{-1}\partial_iH_{(0)}\partial^i\Phi_{(0)}-e^{2\Phi_{(0)}}H_{(0)}^{-2}h_{(0)ti}h_{(0)t}{}^i\,.
\end{eqnarray}
Some of the notation used in these expressions is explained in appendix \ref{eq:appnullsplit}. The way in which we write these coefficients is slightly ambiguous because of the various relations among the coefficients, e.g. we could replace $B_{(2)t}$ by $h_{(2)tt}$ using \eqref{eq:B2t}. In the way we write the coefficients we consider the set of fields: $H_{(0)}$, $h_{(0)it}$, $h_{(2)tt}$, $\Pi_{(0)ij}$, $\Phi_{(0)}$, $\Phi_{(2)}$, $B_{(2)i}$, $B_{(4)t}$, $h_{(6)tt}$, $h_{(4)ti}$, $h_{(4)ij}$, $\phi_{(4)}$ and $B_{(2)r}$ as arbitrary boundary functions whose specification fix the asymptotic expansion. The corresponding 5-dimensional data is given by the set: $\chi_{(0)}$, $\chi_{(4)}$, $\phi_{(0)}$, $\phi_{(4)}$, $H_{(0)}$, $N_{(0)}^i$, $\Pi_{(0)ij}$, $\hat t_{\hat a\hat b}$. The data in $\hat t_{\hat a\hat b}$ is constrained by \eqref{eq:tracet} and \eqref{eq:divt}. Since the reduction distributes the components of $\hat t_{\hat a\hat b}$ over the functions $\Phi_{(2)}$ ($\hat t_{uu}$), $B_{(2)i}$ ($\hat t_{iu}$), $B_{(4)t}$ ($\hat t_{tu}$), $h_{(6)tt}$ ($\hat t_{tt}$), $h_{(4)ti}$ ($\hat t_{ti}$) and $h_{(4)ij}$ ($\hat t_{ij}$) these functions must satisfy the constraints that result from reducing \eqref{eq:tracet} and \eqref{eq:divt}.

The transition from the variables $(A_\mu,\chi)$ to $B_\mu$ via \eqref{eq:B} at the level of the expansions is much more straightforward when $\chi_{(0)}$ is a constant. This can be seen from \eqref{eq:B0t} to \eqref{eq:B0r} or from the fact that in that case \eqref{eq:Nsquaredtoh2tt} becomes a relation among the coefficients with the left hand side equal to $h_{(0)ti}h_{(0)t}{}^i$.  

\subsection{Radial gauge in Einstein frame}

In the frame \eqref{eq:metricdecomposition5Dto4D} the expansions \eqref{eq:htt} to \eqref{eq:phi} form an asymptotically locally Lifshitz space-time according to the definition of \cite{Ross:2011gu}. We will now discuss the expansion from the point of view of the 4-dimensional Einstein frame to see in which sense also in that case we are dealing with an AlLif space-time. In Einstein frame the metric, using the above expansions, takes the following form
\begin{eqnarray}
g^{\text{E}}_{rr} & = & \frac{1}{r^2}e^{\Phi_{(0)}}+\mathcal{O}(\log r)\,,\\
g^{\text{E}}_{tt} & = & \frac{1}{r^4}e^{\Phi_{(0)}}h_{(0)tt}+\mathcal{O}\left(\frac{\log r}{r^2}\right)\,,\\
g^{\text{E}}_{ti} & = & \frac{1}{r^2}e^{\Phi_{(0)}}h_{(0)ti}+\mathcal{O}(\log r)\,,\\
g^{\text{E}}_{ij} & = & \frac{1}{r^2}e^{\Phi_{(0)}}h_{(0)ij}+\mathcal{O}(\log r)\,.
\end{eqnarray}
To write down a Fefferman--Graham type expansion in Einstein frame requires that we write $g_{\mu\nu}^{\text{E}}$ in radial gauge. To this end consider the metric $g^{\text{E}}_{\mu\nu}$ and consider coordinate transforming it,
\begin{equation}
{g'}^{\text{E}}_{\!\!\mu\nu}(x)=g^{\text{E}}_{\mu\nu}(x)+\delta g^{\text{E}}_{\mu\nu}(x)=g^{\text{E}}_{\mu\nu}(x)+\mathcal{L}_{\xi}g^{\text{E}}_{\mu\nu}(x)\,,
\end{equation}
where the coordinates transform as ${x'}^{\mu}=x^{\mu}-\xi^{\mu}$. We require that
\begin{eqnarray}
\delta g^{\text{E}}_{ra}=\mathcal{L}_{\xi}g^{\text{E}}_{ra} & = & g^{\text{E}}_{ab}\partial_r\xi^b+g_{rr}^{\text{E}}\partial_a\xi^r=0\,,\label{eq:deltagra}\\
\delta g^{\text{E}}_{rr}=\mathcal{L}_{\xi}g^{\text{E}}_{rr} & = & 2\left(\partial_r\xi^r-\frac{1}{r}\xi^r+\frac{1}{2}\xi^\mu\partial_\mu\Phi\right)g_{rr}^{\text{E}}=-\left(\xi^\mu\partial_\mu\Phi\right)g^{\text{E}}_{rr}\,,\label{eq:deltagrr}
\end{eqnarray}
where $\delta\Phi=\xi^\mu\partial_\mu\Phi$. This is the infinitesimal version of a coordinate transformation that brings us to radial gauge in Einstein frame. The general solution to the condition \eqref{eq:deltagra} is given by 
\begin{equation}
\xi^a = \xi^a_{(0)}(x)-\int\frac{dr}{r^2}h^{ab}\partial_b\xi^r\,.\label{eq:xia}
\end{equation}
The equation \eqref{eq:deltagrr} can be written as
\begin{equation}\label{eq:defxir}
\partial_r\xi^r-\frac{1}{r}\xi^r+\xi^r\partial_r\Phi+\xi^a_{(0)}\partial_a\Phi-\partial_a\Phi\int\frac{dr}{r^2}h^{ab}\partial_b\xi^r=0\,.
\end{equation}
Using the expansions for $\Phi$ and $h^{ab}$ given in \eqref{eq:phi}, \eqref{eq:invhtt} to \eqref{eq:invhij} we can deduce that at leading order the equation for $\xi^r$ simplifies to
\begin{equation}
\partial_r\xi^r-\frac{1}{r}\xi^r+\xi_{(0)}^a\partial_a\Phi_{(0)}=0\,.
\end{equation}
The solution to this equation is given by
\begin{equation}
\xi^r  =r\left(\xi^r_{(0)}(x)+\log r\xi^r_{(0,1)}(x)\right)\,,\label{eq:xir}
\end{equation}
where $\xi^r_{(0,1)}=-\xi_{(0)}^a\partial_a\Phi_{(0)}$. As a solution to \eqref{eq:defxir} the error in \eqref{eq:xir} is $\mathcal{O}(r^3(\log r)^2)$.

We note that when $\Phi_{(0)}$ is constant then at leading order the metric $g^{\text{E}}_{\mu\nu}=e^{\Phi}g_{\mu\nu}$ with $g_{\mu\nu}$ as given via \eqref{eq:metricdecomposition5Dto4D} agrees with a radial gauge coordinate system with a Lifshitz length scale that is given by $e^{\Phi_{(0)}}$ measured in units of the AdS length scale which we set equal to one.

Knowing the coordinate transformation at leading order in $\xi^r$ is good enough to decide whether we obtain an AlLif space-time for general boundary dependence of $\phi_{(0)}$ and thus $\Phi_{(0)}$ by looking at the leading terms in the FG expansion in radial gauge in Einstein frame. Solving for $\xi^r$ beyond leading order would not modify the leading behavior of the metric but only affect it at subleading orders. 

We now use equations \eqref{eq:xia} and \eqref{eq:xir} to work out the effect of the coordinate transformation on $h_{ab}$ to radial gauge at leading order using
\begin{equation}\label{eq:deltahab}
\delta h_{ab}=\xi^{c}\partial_{c}h_{ab}+h_{ac}\partial_{b}\xi^{c}+h_{cb}\partial_{a}\xi^{c}+\xi^r\partial_r h_{ab}\,.
\end{equation}

Using \eqref{eq:xia} and \eqref{eq:xir} we see that due to the $r\log r$ term in \eqref{eq:xir} we get via the last term in \eqref{eq:deltahab} logarithmic violations of the leading Lifshitz behavior. For example the leading term in $\delta h_{tt}$ is of order $r^{-4}\log r$ whereas for pure Lifshitz we only have $r^{-4}$. The $r^{-4}\log r$ term disappears if and only if we take $\Phi_{(0)}$ constant. 

It is nonetheless useful to perform the analysis of holographic renormalization for arbitrary $\Phi_{(0)}$ because it allows us to treat $\Phi_{(0)}$ as a source and compute the vev for the dual operator. Further turning on $\Phi_{(0)}$ as a non-constant boundary field may be an interesting class of deformations in their own right. It would be interesting to study this more precisely from a renormalization group point of view. 

\subsection{Asymptotically locally $z=2$ Lifshitz space-times}\label{subsec:AlLif}

We are now in a position to define (from a 5-dimensional perspective) the notion of an AlLif space-time. We will call a solution to the equations of motion \eqref{eq:4DEinsteineqs} to \eqref{eq:Aeom} AlLif if and only if the 5-dimensional uplift of this solution (which always exists as the reduction is consistent) satisfies the following properties
\begin{eqnarray}
\hspace{-.7cm}\hat\phi_{(0)}\!\! & = &\!\! \text{cst}\,,\\
\hspace{-.7cm}\hat\chi_{(0)}\!\! & = &\!\! ku+\chi_{(0)}(x)\,,\\
\hspace{-.7cm}\hat h_{(0)\hat a\hat b}\!\! & = &\!\! \text{such that it admits a hypersurface orthogonal null Killing vector $\partial_u$}\,.
\end{eqnarray}
We will show that this agrees nicely with the definition of an AlLif space-time as given in \cite{Ross:2011gu}. When the metric $\hat h_{(0)\hat a\hat b}$ is conformally flat and $\chi_{(0)}$ constant we call the reduced space-time asymptotically Lifshitz.

The reduced solution was already AlLif in the frame defined by \eqref{eq:metricdecomposition5Dto4D}. Now that we require $\Phi_{(0)}$ to be constant it is also guaranteed to be AlLif in Einstein frame. To compare with the Vielbein based definition of AlLif space-times as given in \cite{Ross:2011gu} we can simply decompose the metric $g_{\mu\nu}$ into Vielbeins. Doing so we obtain
\begin{eqnarray}
e^{\underline{t}} & = & r^{-2}\tilde e^{\underline{t}}_t dt+\tilde e^{\underline{t}}_i dx^i\,,\\
e^{\underline{i}} & = & r^{-1}\tilde e^{\underline{i}}_t dt+r^{-1}\tilde e^{\underline{i}}_i dx^i\,,
\end{eqnarray}
where the tangent space metric $\eta_{\underline{a}\underline{b}}$ is
\begin{equation}
\eta_{\underline{t}\underline{t}}=-1\,,\qquad\eta_{\underline{t}\underline{i}}=0\,,\qquad\eta_{\underline{i}\underline{j}}=\delta_{\underline{i}\underline{j}}\,.
\end{equation}
The boundary conditions are such that
\begin{equation}
\tilde e^{\underline{t}}_t\vert_{r=0}\,,\qquad\tilde e^{\underline{i}}_i\vert_{r=0}\,,
\end{equation}
are nonzero functions of the boundary coordinates whereas
\begin{equation}
\tilde e^{\underline{t}}_i\vert_{r=0}\,,\qquad\tilde e^{\underline{i}}_t\vert_{r=0}\,,
\end{equation}
can be chosen freely (zero or nonzero functions of the boundary coordinates). These boundary conditions nicely agree with those of \cite{Ross:2011gu} including the condition that $r^2 e^{\underline{t}}$ is hypersurface orthogonal as $r$ goes to zero. This is tied to the fact that we have chosen coordinates such that $\hat h_{(0)iu}=0$ which in turn is related to choosing adapted coordinates for $\hat h_{(0)\hat a\hat b}$ to make the hypersurface orthogonality of the null Killing vector $\partial_u$ manifest.

\section{Lifshitz counterterms and scale anomalies}\label{sec:anomalies}

With the results of the previous two sections we are now in a position to discuss the counterterms for the AlLif space-times, i.e. to work out the form of $S_{\text{ct}}$ in \eqref{eq:4DactionB} and to work out the anomaly counterterms on-shell. From the 5-dimensional point of view the on-shell anomaly counterterm is related to the trace anomaly \eqref{eq:tracet}. Upon Scherk--Schwarz dimensional reduction we will see that from the 4-dimensional perspective we are dealing with anisotropic rescalings and two associated anomaly terms, one second order and one fourth order in derivatives.

\subsection{Anisotropic conformal rescalings}

Conformal rescalings of the boundary metric $\hat h_{(0)\hat a\hat b}$ can be generated by Penrose--Brown--Henneaux (PBH) transformations \cite{Penrose:1986ca,Brown:1986nw}, i.e. diffeomorphisms that preserve the radial gauge choice. Infinitesimally these transformations act on the 5-dimensional fields as
\begin{eqnarray}
\delta\hat g_{\hat\mu\hat\nu} & = & \mathcal{L}_{\hat\xi}\hat g_{\hat\mu\hat\nu}\,,\\
\delta\hat\phi & = & \mathcal{L}_{\hat\xi}\hat\phi\,,\\
\delta\hat\chi & = & \mathcal{L}_{\hat\xi}\hat\chi\,,
\end{eqnarray}
such that $\mathcal{L}_{\hat\xi}\hat g_{rr}=\mathcal{L}_{\hat\xi}\hat g_{r\hat a}=0$ so that the radial gauge of the 5-dimensional metric \eqref{eq: sol metric gauge} is preserved. The solution to these equations gives
\begin{eqnarray}
\hat\xi^r & = & r\hat\xi^r_{(0)}\,,\\
\hat\xi^a & = & \hat\xi^{\hat a}_{(0)}-\int\frac{dr}{r} \hat h^{\hat a\hat b}\partial_{\hat b}\hat\xi_{(0)}^r\,,
\end{eqnarray}
where $\hat\xi^r_{(0)}$ and $\hat\xi^{\hat a}_{(0)}$ are independent of $r$. Acting with such diffeomorphisms assuming $\hat\xi_{(0)}^r\neq 0$ on the 5-dimensional solution leads to conformal rescalings and reparametrizations of the boundary metric $\hat h_{(0)\hat a\hat b}$ via
\begin{equation}
\delta\hat h_{\hat a\hat b}=\hat\xi^{\hat c}\partial_{\hat c}\hat h_{\hat a\hat b}+\hat h_{\hat a\hat c}\partial_{\hat b}\hat\xi^{\hat c}+\hat h_{\hat b\hat c}\partial_{\hat a}\hat\xi^{\hat c}+\hat\xi^r\partial_r\hat h_{\hat a\hat b}\,.
\end{equation}
If we further demand that the transformed metric still satisfies the reduction Ansatz then we must also require that $\hat\xi^r_{(0)}$ and $\hat\xi^{\hat a}_{(0)}$ are independent of $u\,$\footnote{A similar restriction for the AdS Penrose--Brown--Henneaux transformations has also been observed in Fefferman--Graham expansions for asymptotically $z=2$ Schr\"odinger space-times that can be obtained from asymptotically AdS space-times via the so-called TsT transformation \cite{Hartong:2010ec}.}. This means that the boundary rescalings and diffeomorphisms preserve the existence of a hypersurface orthogonal null Killing vector given by $\partial_u\,${}\footnote{These restrictions are not strong enough to preserve the form of the parametrization \eqref{eq:bdrymetricparametrization}. That will only be the case if we furthermore demand that $\hat\xi^t_{(0)}$ is independent of $x^i$ and is thus a function of $t$ only.}. 

The finite version of these transformations (with $\hat\xi^{\hat a}_{(0)}=0$) transform the leading terms in the Fefferman--Graham expansion as follows
\begin{eqnarray}
\hat h_{(0)\hat a\hat b} & \rightarrow & \Omega^2\hat h_{(0)\hat a\hat b}\,,\\
\hat\chi_{(0)} & \rightarrow & \hat\chi_{(0)}\,,\\
\hat\phi_{(0)} & \rightarrow & \hat\phi_{(0)}\,,
\end{eqnarray}
with $\partial_u\Omega=0$. In the parametrization \eqref{eq:bdrymetricparametrization} the conformal rescalings act as
\begin{eqnarray}
H_{(0)} & \rightarrow & \Omega^2H_{(0)}\,,\label{eq:weyl1}\\
\Pi_{(0)ij} & \rightarrow & \Omega^2\Pi_{(0)ij}\,,\\
N^i_{(0)} & \rightarrow & \Omega^{-2}N_{(0)}^i\,.\label{eq:weyl3}
\end{eqnarray}
Further the scalars $\Phi_{(0)}$, $\phi_{(0)}$ and $\chi_{(0)}$ transform with weight zero. This implies that $h_{(0)tt}$ scales with weight 4 while $h_{(0)ti}$ and $h_{(0)ij}$ scale with weight two. These are precisely the anisotropic conformal rescalings of \cite{Horava:2009vy}. We will now study the associated anisotropic conformal anomalies by dimensional reduction of the 5-dimensional counterterm action.

\subsection{Dimensional reduction of the counterterm action}

Performing a dimensional reduction of the counterterm action \eqref{eq:Sct1} we obtain
\begin{eqnarray}
    S_{\text{ct}} & = & \frac{2\pi L}{\kappa^2_5}
    \int_{\partial\mathcal{M}}d^3x\sqrt{-h}e^{\Phi}\left[-3
    -\frac{1}{4}\left(R_{(h)} - \frac{1}{4}e^{2 \Phi} F^2
    - \frac{1}{2} (\partial\phi)^2
    - \frac{k^2}{2}e^{2 \phi}B^2
    \right.\right.\nonumber\\
    &&\left.\left.- \frac{k^2}{2}e^{2 \phi- 2 \Phi} \right)\right]+(\log r)\frac{2\pi L}{\kappa^2_5}
    \int_{\partial\mathcal{M}}d^3x\sqrt{-h}e^{\Phi}\left(\mathcal{A}^{(0)}+\mathcal{A}^{(2)}+\mathcal{A}^{(4)}\right)\,,\label{eq:reduced ct action}
\end{eqnarray}
where
\begin{eqnarray}
\mathcal{A}^{(0)} & = & \frac{k^4}{12}e^{4\phi}\left(B^2+e^{-2\Phi}\right)^2\,,
\end{eqnarray}
\begin{eqnarray}
\mathcal{A}^{(2)} & = & -\frac{k^2}{8}e^{2\phi}B^aB^b\left(R_{(h)ab} - \nabla^{(h)}_a\partial_b\Phi
        - \partial_a\Phi\partial_b\Phi
        - \frac{1}{2}e^{2 \Phi}F_a{}^c F_{bc}
        - \frac{1}{2}\partial_a\phi\partial_b\phi\right)\nonumber\\
        &&+\frac{k^2}{8}e^{2\phi}B^a\left(\nabla^{(h)b} F_{ab}
                +3F_{ab}\partial^b\Phi\right)+\frac{k^2}{8}e^{2\phi-2\Phi}\left(\square^{(h)}\Phi + (\partial\Phi)^2
        - \frac{1}{4}e^{2 \Phi} F^2\right)\nonumber\\
        &&+\frac{k^2}{24}e^{2\phi}\left(B^2+e^{-2\Phi}\right)\left(R_{(h)}
        - 2 \square^{(h)}\Phi - 2 (\partial\Phi)^2
        - \frac{1}{4}e^{2 \Phi} F^2 - \frac{1}{2} (\partial\phi)^2\right.\nonumber\\
        &&\left.-3\left(\square^{(h)}\phi +
\partial^a\phi\partial_a\Phi\right)\right)+\frac{k^2}{16}e^{2\phi}\left(\nabla^{(h)}_aB^a+B^a\partial_a\Phi+2B^a\partial_a\phi\right)^2\,,\\
\mathcal{A}^{(4)} & = & \frac{1}{8}
        \left(R_{(h)ab} - \nabla^{(h)}_a\partial_b\Phi
        - \partial_a\Phi\partial_b\Phi
        - \frac{1}{2}e^{2 \Phi}F_a{}^c F_{bc}
        - \frac{1}{2}\partial_a\phi\partial_b\phi
        \right)^2\nonumber\\
&&+\frac{1}{16}e^{2\Phi}\left(\nabla^{(h)b}F_{ab}+3F_{ab}\partial^b\Phi\right)^2+\frac{1}{8}\left(\square^{(h)}\Phi + (\partial\Phi)^2
        - \frac{1}{4}e^{2 \Phi} F^2\right)^2\nonumber\\
&&-\frac{1}{24}\left(R_{(h)}
        - 2 \square^{(h)}\Phi - 2 (\partial\Phi)^2
        - \frac{1}{4}e^{2 \Phi} F^2 - \frac{1}{2} (\partial\phi)^2 \right)^2\nonumber\\
        &&+\frac{1}{16}\left(\square^{(h)}\phi+\partial^a\phi\partial_a\Phi\right)^2\,.
\end{eqnarray}
The superscript on $\mathcal{A}$ refers to the number of derivatives. 

The local counterterms given in the first line of \eqref{eq:reduced ct action} agree exactly with the counterterms given in \cite{Mann:2011hg} for what they call the minimal action provided we set $\Phi$ and $\phi$ equal to constants such that $\Phi-\phi=\log\frac{k}{2}$. To compare with the expression given \cite{Mann:2011hg} one must perform some mild field redefinitions. Even though setting $\Phi$ and $\phi$ equal to constants is not a consistent truncation from the model discussed here to the massive vector model without any scalars it is interesting that we nonetheless get the same answer. This is because it is consistent and in fact necessary in order to get AlLif solutions to set $\Phi_{(0)}$ and $\phi_{(0)}$ equal to constants, i.e. the scalars become asymptotically constant. Since in this case the counterterms do not depend on the $r$ dependent part of $\Phi$ and $\phi$ (from a 5-dimensional point of view this is to say that the divergent parts of the on-shell action do not depend on $\hat\phi_{(4)}$ and $\hat t_{\hat a\hat b}$) we should get the same answer as for the massive vector model without scalars. Since for asymptotically constant scalars the term in $S_{\text{ct}}$ proportional to $B^2+e^{-2\Phi}$ in the first line of \eqref{eq:reduced ct action} is at least of order $r^0$, i.e. at best a finite counterterm (we did not check if the coefficient is nonzero), the result also agrees with the local counterterm used in \cite{Ross:2011gu,Griffin:2011xs} for the no scalar massive vector model. 

Even though the on-shell 4-dimensional action, \eqref{eq:4Daction} and \eqref{eq:reduced ct action}, is finite by construction we have checked that it is finite when we evaluate it for the reduced expansions of appendix \ref{app:4DexpansionsAandchi}. For this purpose one needs to expand the fields up to the orders indicated in \eqref{eq:httapp} to \eqref{eq:chiapp}. We have not listed all coefficients for reasons as explained just below \eqref{eq:phi2}. The check of the finiteness has been performed with the software package Cadabra \cite{Peeters:2006kp,Peeters:2007wn}. We consider this an important check on our algebra. The form of the counterterms is not unique (and this has nothing to do with the freedom to add finite counterterms). There are many ways of rewriting them that would equally lead to a finite renormalized on-shell action with the same on-shell expression. We have simply chosen a form that one obtains from the reduction (modulo a few total derivatives in the boundary Lagrangian that have been removed).

We will now use this result to study the Lifshitz scale anomaly by evaluating the term proportional to $\log r$ in $S_{ct}$ on-shell. When doing so we will make one simplifying assumption which is to take the 4-dimensional $\chi_{(0)}$ constant. For $\chi_{(0)}$ constant we find that $\mathcal{A}^{(0)}$ is zero. Otherwise it would have been second order in derivatives of $\chi_{(0)}$. Further we have, using that $\Phi_{(0)}-\phi_{(0)}=\log\frac{k}{2}$ and that $\phi_{(0)}$ is constant,
\begin{eqnarray}
\hspace{-.7cm}\frac{2\pi L}{\kappa_5^2}\int_{\partial\mathcal{M}}\!\!\!\!d^3x\sqrt{-h}e^{\Phi}\mathcal{A}^{(2)}\!\!\! & = &\!\!\! \frac{2\pi L}{8\kappa_5^2}e^{2\Phi_{(0)}}\!\!\int_{\partial\mathcal{M}}\!\!\!\!dtd^2xH'_{(0)}\sqrt{\Pi_{(0)}}\!\left(4K_{(0)ij}K_{(0)}^{ij}-2K_{(0)}^2\right)\,,\label{eq:anomaly2}\\
\hspace{-.7cm}\frac{2\pi L}{\kappa_5^2}\int_{\partial\mathcal{M}}\!\!\!\!d^3x\sqrt{-h}e^{\Phi}\mathcal{A}^{(4)}\!\!\! & = &\!\!\!\frac{2\pi L}{48\kappa_5^2}e^{2\Phi_{(0)}}\!\!\int_{\partial\mathcal{M}}\!\!\!\!dtd^2xH'_{(0)}\sqrt{\Pi_{(0)}}\!\left(\mathcal{R}_{(0)}+D^{(0)i}\partial_i\log H'_{(0)}\right)^2\,,\nonumber\\
&&\label{eq:anomaly4}
\end{eqnarray}
where we defined
\begin{equation}
H_{(0)}'=(-h_{(0)tt})^{1/2}=H_{(0)}e^{-\Phi_{(0)}}
\end{equation}
and where 
\begin{equation}
K_{(0)ij}=\frac{1}{2H'_{(0)}}\left(\partial_t\Pi_{(0)ij}-D^{(0)}_ih_{(0)tj}-D^{(0)}_jh_{(0)ti}\right)\,,\qquad K_{(0)}=\Pi_{(0)}^{ij}K_{(0)ij}\,,
\end{equation}
is the extrinsic curvature and its trace. The integrands in \eqref{eq:anomaly2} and \eqref{eq:anomaly4} are invariant under the anisotropic Weyl rescalings \eqref{eq:weyl1} to \eqref{eq:weyl3}. The on-shell expression for the anomaly counterterm of \eqref{eq:reduced ct action} forms an action that is of the Horava--Lifshitz type with nonzero potential term for $z=2$ conformal gravity \cite{Horava:2008ih,Griffin:2011xs}.

The expression for the anomaly at second order in derivatives \eqref{eq:anomaly2} agrees with what has been found in \cite{Griffin:2011xs,Baggio:2011ha}. The anomaly at fourth order in derivatives \eqref{eq:anomaly4} has been shown in \cite{Griffin:2011xs,Baggio:2011ha} to exist on general grounds but was not observed in the no scalar massive vector model. Its presence here does not seem to rely on the presence of scalars in the analysis. It would be interesting to understand this better. The term in parenthesis in \eqref{eq:anomaly2} and \eqref{eq:anomaly4} vanishes for asymptotically Lifshitz space-times, i.e. for a conformally flat boundary metric $\hat h_{(0)\hat a\hat b}$. In the notation of \cite{Baggio:2011ha} we have the following values for the central charges $C_1$, $C_2${}\footnote{We thank the authors of \cite{Baggio:2011ha} for useful discussions.},
\begin{eqnarray}
C_1 & = & \frac{2\pi L}{64\pi G_5}l^2e^{2\Phi_{(0)}}\,,\\
C_2 & = & \frac{2\pi L}{384\pi G_5}l^2e^{2\Phi_{(0)}}=\frac{1}{6}C_1\,,
\end{eqnarray}
where we have inserted the AdS length parameter $l$ of the 5-dimensional asymptotically locally AdS space-times.
This can also be written as
\begin{equation}
C_1=\frac{l^2_{\text{Lif}}}{64\pi G_4}=6C_2\,,
\end{equation}
where
\begin{equation}
l^2_{\text{Lif}}=l^2e^{\Phi_{(0)}}
\end{equation}
is the Lifshitz length parameter\footnote{In Einstein frame the coefficient of the $\tfrac{dr^2}{r^2}$ term in metric \eqref{eq:Lifshitzmetric} is given by $l^2e^{\Phi_{(0)}}$.} and where $G_4$ is the 4-dimensional Newton's constant given by
\begin{equation}
\frac{1}{G_{4}}=\frac{2\pi L e^{\Phi_{(0)}}}{G_{5}}
\end{equation}
in which $L e^{\Phi_{(0)}}$ is the asymptotic value of the radius of the compactification circle.

\section{Discussion}

We have performed holographic renormalization for AlLif space-times with $z=2$ in the context of solutions of type IIB supergravity. The approach was based on the observation that a 4-dimensional $z=2$ Lifshitz space-time in IIB string theory can be obtained by combining a stack of extremal D3-branes with an axion plane wave. From a 5-dimensional point of view the intersection of the D3-brane and the axion wave leads to a $z=0$ Schr\"odinger space-time which is an asymptotically AdS space-time. The relation to a $z=2$ Lifshitz space-time is then via Scherk--Schwarz reduction. This situation has been observed, in various forms, in \cite{Balasubramanian:2010uk,Donos:2010tu,Cassani:2011sv,Costa:2010cn,Chemissany:2011mb}.

As mentioned in \cite{Costa:2010cn} the reduction from the point of view of the boundary theory is along a lightlike circle and should therefore be viewed as some DLCQ of $\mathcal{N}=4$ SYM in the background of a theta angle that depends linearly on the null circle coordinate leading to Lifshitz Chern--Simons gauge theory \cite{Balasubramanian:2011ua}. This fact however does not prevent us from performing holographic renormalization in the bulk as the reduction in the bulk is everywhere along a spacelike circle.

From a 5-dimensional point of view the boundary of the AlAdS space-time must admit a hypersurface orthogonal null Killing vector $\partial_u$. This vector $\partial_u$ generates the compact null circle on the boundary. We have used boundary coordinates \eqref{eq:bdrymetricparametrization} that are suitably adapted to the existence of such a vector field and this has played a central role in the construction of AlLif space-times. It would be interesting to define the boundary conditions for having an AlLif space-time in a coordinate independent manner. Once this parametrization has been chosen, the structure of the 4-dimensional Fefferman--Graham expansions agrees with the boundary condition for AlLif space-times given in \cite{Ross:2011gu} provided we choose the 5-dimensional dilaton to asymptote to a constant. Regarding the 5-dimensional axion there is the restriction that it asymptotes to $\hat\chi_{(0)}=ku+\chi_{(0)}(x)$. From a 4-dimensional perspective one then has the possibility to describe the boundary data from the point of view of either the $(A_\mu,\chi)$ (with a gauge symmetry) or the $B_\mu$ variables. When the 4-dimensional $\chi_{(0)}$ is non-constant the relation between these two sets of variables from the boundary point of view, i.e. the free functions appearing in the FG expansions, is not so simple. It would be interesting to understand this better and to see if there is a preferred set of variables. 

Upon dimensional reduction of the local counterterms of the AlAdS space-time we obtain the local counterterms of the AlLif space-times and these nicely agree with what has been found in the literature so far \cite{Mann:2011hg,Ross:2011gu,Griffin:2011xs,Baggio:2011ha}. Further, upon dimensional reduction of the anomaly counterterms we find that the 4-dimensional anomaly counterterm evaluated on-shell (for AlLif space-times with $\chi_{(0)}$ constant) consists of two pieces that together form the action of $z=2$ conformal gravity in $2+1$ dimensions with nonzero potential term \cite{Horava:2008ih,Griffin:2011xs}. The presence of the potential term has so far not been seen in studies of the no scalar massive vector model. At the same time in the setting in which we computed the on-shell anomaly all scalars where asymptotically constant and our on-shell anomalies do not depend on the scalars. It would be interesting to understand this situation better. What is noteworthy about the reduced on-shell anomaly is that there now appear two central charges (in the notation of \cite{Baggio:2011ha} these are $C_1$ and $C_2$) that are proportional to each other. From the reduction we can see that both originate from the single central charge in 5-dimensions. It would be interesting to understand this from the dual field theory point of view, i.e. the DLCQ of $\mathcal{N}=4$ SYM in the background of a theta angle that depends linearly on the null circle coordinate. 

Further it would also be of interest to see if this setup can be used to understand better the asymptotic symmetry group for AlLif space-times (see \cite{Mann:2011hg}) and to see if it is possible to understand the presence of the two central charges from that point of view. In the 5-dimensional case the central charge shows up in the transformation of the boundary stress tensor under a PBH transformation. It would interesting to investigate if there are similar statements possible for AlLif space-times and what the role of the stress tensor complex of \cite{Ross:2009ar,Ross:2011gu} is in this respect.

\section*{Acknowledgements}

Many calculations have been checked using the program Cadabra \cite{Peeters:2006kp,Peeters:2007wn}. We express our gratitude to Matthias Blau, Jakob Gath, Ulf Gran, Ricardo Monteiro, Niels Obers, Balt van Rees, Bert Vercnocke and Amitabh Virmani for useful discussions and correspondences. W.C. is supported in part by the Natural Sciences and Engineering Research Council (NSERC) of Canada. The work of D.G. and B.R. was supported by the Swiss National Science Foundation. D.G. thanks CQUeST, Seoul, for warm hospitality (during the later stages of this work). The work of J.H. was supported by the Danish National Research Foundation project ÒBlack holes and their role in quantum gravityÓ. J.H. also thanks the Isaac Newton Institute, Cambridge, for hospitality during the later stages of this work.

\appendix

\section{Double null split}\label{eq:appnullsplit}

In this appendix we collect expressions for the Christoffel connections and the Ricci tensor components using the following double null split of the boundary metric of the 5-dimensional AlAdS space-times
\begin{equation}
\hat h_{(0)\hat a\hat b}dx^{\hat a}dx^{\hat b}=2H_{(0)}dudt+\Pi_{(0)ij}\left(dx^i+H_{(0)}N_{(0)}^idt\right)\left(dx^j+H_{(0)}N_{(0)}^jdt\right)\,,
\end{equation}
where all metric components are arbitrary functions of $t$ and $x^i$ but do not depend on $u$. The nonzero inverse metric components are given by
\begin{equation}
\hat h_{(0)}^{ut}=H_{(0)}^{-1}\,,\qquad\hat h_{(0)}^{ui}=-N_{(0)}^i\,,\qquad\hat h_{(0)}^{ij}=\Pi_{(0)}^{ij}\,.
\end{equation}
For the determinant we have
\begin{equation}
\sqrt{-\hat h_{(0)}}=H_{(0)}\sqrt{\Pi_{(0)}}\,,
\end{equation}
where $\Pi_{(0)}$ is the determinant of the 2 by 2 metric $\Pi_{(0)ij}$. The nonzero Christoffel connections are
\begin{eqnarray}
\hat\Gamma^{(0)u}_{ut} & = & \frac{1}{2}N_{(0)}^i\partial_i H_{(0)}\,,\\
\hat\Gamma^{(0)u}_{ui} & = & \frac{1}{2}H_{(0)}^{-1}\partial_i H_{(0)}\,,\\
\hat\Gamma^{(0)u}_{tt} & = & -\frac{1}{2}H_{(0)}N_{(0)}^iN_{(0)}^j\partial_t\Pi_{(0)ij}+H_{(0)}\left(N_{(0)}^i\partial_i H_{(0)}\right)N_{(0)j}N_{(0)}^j\nonumber\\
&&+H_{(0)}^2N_{(0)}^iN_{(0)}^jD^{(0)}_iN_{(0)j}\,,\\
\hat\Gamma^{(0)u}_{ti} & = & \frac{1}{2}N_{(0)j}N_{(0)}^j\partial_i H_{(0)}+\frac{1}{2}H_{(0)}N_{(0)}^j\left(D^{(0)}_iN_{(0)j}+D^{(0)}_jN_{(0)i}\right)\nonumber\\
&&+\frac{1}{2}N_{(0)i}N_{(0)}^j\partial_jH_{(0)}-\frac{1}{2}N_{(0)}^j\partial_t\Pi_{(0)ij}\,,
\end{eqnarray}
\begin{eqnarray}
\hat\Gamma^{(0)u}_{ij} & = & \frac{1}{2}H_{(0)}^{-1}\left(\partial_i H_{(0)}\right)N_{(0)j}+\frac{1}{2}H_{(0)}^{-1}\left(\partial_j H_{(0)}\right)N_{(0)i}-\frac{1}{2}H_{(0)}^{-1}\partial_t\Pi_{(0)ij}\nonumber\\
&&+\frac{1}{2}\left(D^{(0)}_iN_{(0)j}+D^{(0)}_jN_{(0)i}\right)\,,\\
\hat\Gamma^{(0)t}_{tt} & = & H_{(0)}^{-1}\partial_t H_{(0)}\,,\\
\hat\Gamma^{(0)t}_{ti} & = & \frac{1}{2}H_{(0)}^{-1}\partial_iH_{(0)}\,,\\
\hat\Gamma^{(0)k}_{ut} & = & -\frac{1}{2}\partial^kH_{(0)}\,,\\
\hat\Gamma^{(0)k}_{tt} & = & H_{(0)}\partial_tN_{(0)}^k+H_{(0)}\Pi^{kl}_{(0)}N_{(0)}^m\partial_t\Pi_{(0)lm}-H_{(0)}N_{(0)i}N_{(0)}^i\partial^k H_{(0)}\nonumber\\
&&-H_{(0)}^2N_{(0)i}D^{(0)k}N_{(0)}^i\,,\\
\hat\Gamma^{(0)k}_{it} & = & \frac{1}{2}H_{(0)}\left(D^{(0)}_iN_{(0)}^k-D^{(0)k}N_{(0)i}\right)-\frac{1}{2}N_{(0)i}\partial^k H_{(0)}+\frac{1}{2}\Pi_{(0)}^{kl}\partial_t\Pi_{(0)il}\,,\\
\hat\Gamma^{(0)k}_{ij} & = & C^{(0)k}_{ij}\equiv\frac{1}{2}\Pi_{(0)}^{kl}\left(\partial_i\Pi_{(0)jl}+\partial_j\Pi_{(0)il}-\partial_l\Pi_{(0)ij}\right)\,,
\end{eqnarray}
where indices are raised and lowered with $\Pi_{(0)ij}$, so $N_{(0)i}=\Pi_{(0)ij}N_{(0)}^j$ not to be confused with the $i$th component of $\hat N_{(0)\hat a}$ in \eqref{eq:HSOadaptedcoordinates}, and where $D^{(0)}_i$ is the covariant derivative with respect to the metric $\Pi_{(0)ij}$ and where finally $C^{(0)k}_{ij}$ are the associated Christoffel connections. For the nonzero components of the Ricci tensor we get
\begin{eqnarray}
\hat R_{(0)ut}\!\! & = &\!\! -\frac{1}{2}D^{(0)i}\partial_i H_{(0)}\,,\\
\hat R_{(0)ij}\!\! & = &\!\! \mathcal{R}_{(0)ij}-H_{(0)}^{-1}D^{(0)}_{i}\partial_j H_{(0)}+\frac{1}{2}H_{(0)}^{-2}\partial_iH_{(0)}\partial_jH_{(0)}\,,\\
\hat R_{(0)it}\!\! & = &\!\! \frac{1}{2}\Pi_{(0)}^{kl}\left[D^{(0)}_l\left(\partial_t\Pi_{(0)ik}\right)-D^{(0)}_i\left(\partial_t\Pi_{(0)kl}\right)\right]+\frac{1}{4}H_{(0)}^{-1}\partial_i H_{(0)}\Pi_{(0)}^{kl}\partial_t\Pi_{(0)kl}\nonumber\\
&&+\frac{1}{2}H_{(0)}D^{(0)k}\left(D^{(0)}_iN_{(0)k}-D^{(0)}_kN_{(0)i}\right)-\frac{1}{2}N_{(0)i}D^{(0)k}\partial_k H_{(0)}\nonumber\\
&&+\left(\partial^k H_{(0)}\right)\left(D^{(0)}_iN_{(0)k}-D^{(0)}_kN_{(0)i}\right)-\frac{1}{2}\partial_t\partial_i\log H_{(0)}\,,\\
\hat R_{(0)tt}\!\! & = &\!\! -\frac{1}{2}H_{(0)}^2\left(D^{(0)l}N_{(0)}^i\right)D^{(0)}_lN_{(0)i}-\frac{1}{2}H_{(0)}^2\left(D^{(0)l}N_{(0)}^i\right)D^{(0)}_iN_{(0)l}\nonumber\\
&&-H_{(0)}^2N_{(0)}^iD^{(0)l}D_{(0)l}N_{(0)i}-\frac{1}{2}N_{(0)i}N_{(0)}^i\left(\partial^k H_{(0)}\right)\partial_k H_{(0)}\nonumber\\
&&-H_{(0)}N_{(0)i}N_{(0)}^iD^{(0)j}\partial_j H_{(0)}-3H_{(0)}\left(\partial^k H_{(0)}\right)N_{(0)}^iD^{(0)}_kN_{(0)i}\nonumber\\
&&+\frac{1}{4}\Pi_{(0)}^{km}\Pi_{(0)}^{ln}\left(\partial_t\Pi_{(0)kn}\right)\partial_t\Pi_{(0)lm}+\frac{1}{2}\hat\Pi_{(0)}^{kl}\left(\partial_t\hat\Pi_{(0)kl}\right)H_{(0)}^{-1}\partial_t H_{(0)}\nonumber\\
&&-\frac{1}{2}\Pi^{kl}_{(0)}\partial^2_t\Pi_{(0)kl}+H_{(0)}D^{(0)k}\partial_tN_{(0)k}+\left(\partial^k H_{(0)}\right)\partial_tN_{(0)k}\,.
\end{eqnarray}
One can also define the shift vector $K_{(0)}^i=H_{(0)}N_{(0)}^i$ but it does not make the expressions shorter. The Ricci scalar is given by
\begin{equation}
\hat R_{(0)}=\mathcal{R}_{(0)}-2H_{(0)}^{-1}D^{(0)i}\partial_i H_{(0)}+\frac{1}{2}H_{(0)}^{-2}\partial_i H_{(0)}\partial^i H_{(0)}\,,
\end{equation}
where $\mathcal{R}_{(0)}$ is the Ricci scalar associated with the metric $\Pi_{(0)ij}$.

\section{The 4-dimensional Fefferman--Graham expansions in terms of $A_\mu$ and $\chi$}\label{app:4DexpansionsAandchi}

Here we will present the 4-dimensional Fefferman--Graham expansions for the on-shell configurations of the theory described by the action \eqref{eq:4Daction} in terms of the vector field $A_\mu$ and the axion $\chi$. The reduction Ansatz reads
\begin{eqnarray}
g_{rr} & = & \frac{1}{r^2}\,,\\
g_{ra} & = & 0\,,\\
h_{ab} & = & \hat h_{ab}-\frac{\hat h_{au}\hat h_{bu}}{\hat h_{uu}}\,,\label{eq:habapp}\\
A_r & = & 0\,,\\
A_a & = & \frac{\hat h_{au}}{\hat h_{uu}}\,,\label{eq:Aa}\\
\Phi & = & \frac{1}{2}\log\hat h_{uu}\,,\label{eq:Phiapp}
\end{eqnarray}
Using these equations together with the reduction Ans\"atze \eqref{eq: decomposition of chi} and \eqref{eq: decomposition of phi} as well as the 5-dimensional expansions \eqref{eq: sol metric} to \eqref{eq: sol chi} and the boundary parametrization \eqref{eq:bdrymetricparametrization}, we obtain expansions for the 4-dimensional fields in terms of $H_{(0)}$, $N_{(0)}^i$, $\Pi_{(0)ij}$,
$\phi_{(0)}$, $\chi_{(0)}$ as well as the free parameters appearing at higher order in the expansions \eqref{eq: sol metric} to \eqref{eq: sol chi}, i.e. the reduced versions of the coefficients $\hat t_{\hat a\hat b}$, $\hat\phi_{(4)}$ and $\hat\chi_{(4)}$. These expansions take the following form
\begin{eqnarray}
h_{tt} & = & \frac{1}{r^4}\left(h_{(0)tt} +r^2\log r h_{(2,1)tt}+
r^2h_{(2)tt}+r^4(\log r)^2 h_{(4,2)tt}\right.\nonumber\\
&&\left.+r^4\log r h_{(4,1)tt}+r^4h_{(4)tt}+\mathcal{O}(r^6(\log r)^3)\right)\,,\label{eq:httapp}\\
h_{ti} & = & \frac{1}{r^2}\left(h_{(0)ti}+r^2\log r h_{(2,1)ti}+r^2h_{(2)ti}+\mathcal{O}(r^4(\log r)^2)\right)\,,\label{eq:hitapp}\\
h_{ij} & = & \frac{1}{r^2}\left(h_{(0)ij}+r^2h_{(2)ij}+r^4\log r h_{(4,1)ij}+r^4h_{(4)ij}+\mathcal{O}(r^6(\log r)^2)\right)\,,\label{eq:hijapp}\\
A_t  & = &  \frac{1}{r^2}\left(A_{(0)t} +r^2\log r A_{(2,1)t}+r^2A_{(2)t}+r^4(\log r)^2 A_{(4,2)t}+r^4\log r A_{(4,1)t}\right.\nonumber\\
&&\left.+r^4A_{(4)t}+\mathcal{O}(r^6(\log r)^3)\right)\,,\\
A_i  & = &  A_{(0)i}+r^2\log r A_{(2,1)i}+r^2A_{(2)i}+\mathcal{O}(r^4(\log r)^2)\,,\\
\Phi &  = & \Phi_{(0)} + r^2\log r\Phi_{(2,1)}+ r^2\Phi_{(2)} +r^4(\log r)^2\Phi_{(4,2)}+r^4\log r\Phi_{(4,1)}\nonumber\\
&&+r^4\Phi_{(4)}+\mathcal{O}(r^6(\log r)^3)\,,\\
\phi & = & \phi_{(0)}+r^2\phi_{(2)}+r^4\log r\phi_{(4,1)}+r^4\phi_{(4)}+\mathcal{O}(r^6\log r)\,,\label{eq:phiapp}\\
\chi & = & \chi_{(0)}+r^2\chi_{(2)}+r^4\log r\chi_{(4,1)}+r^4\chi_{(4)}+\mathcal{O}(r^6\log r)\,.\label{eq:chiapp}
\end{eqnarray}

Just as in the massive vector case discussed in section \ref{sec:SSreduction} we have the constraints at leading order
\begin{eqnarray}
h_{(0)tt} & = & -(A_{(0)t})^2e^{2\Phi_{(0)}}\,,\label{eq:constraint0thorderapp}\\
\Phi_{(0)} & = & \phi_{(0)}+\log\frac{k}{2}\,.
\end{eqnarray}
Again we will deal with this by writing
\begin{eqnarray}
A_{(0)t} & = & H_{(0)}e^{-2\Phi_{(0)}}\,,\\
h_{(0)tt} & = & -H_{(0)}^2e^{-2\Phi_{(0)}}\,,
\end{eqnarray}
and eliminating $\phi_{(0)}$ in favor of $\Phi_{(0)}$.
Since we have the following relations
\begin{eqnarray}
h_{(0)ti} & = & H_{(0)}\left(N_{(0)i}-\frac{1}{k}\partial_i\chi_{(0)}\right)\,,\\
h_{(0)ij} & = & \Pi_{(0)ij}\,,
\end{eqnarray}
we will keep writing $\Pi_{(0)ij}$ and replace $N_{(0)i}$ by $H_{(0)}^{-1}h_{(0)ti}+\frac{1}{k}\partial_i\chi_{(0)}$ in order to express the 4D expansions in terms of the coefficients appearing in \eqref{eq:httapp} to \eqref{eq:chiapp}. Translating in this manner the reduced expansions we obtain the following coefficients
\begin{eqnarray}
h_{(2,1)tt}\!\! & = &\!\! 2H_{(0)}^2e^{-2\Phi_{(0)}}\Phi_{(2,1)}\,,\\
h_{(2)tt}\!\! & = &\!\! h_{(0)ti}h_{(0)t}{}^i+\frac{2}{k}H_{(0)}h_{(0)ti}\partial^i\chi_{(0)}+\frac{1}{k^2}H_{(0)}^2\partial_i\chi_{(0)}\partial^i\chi_{(0)}-2H_{(0)}A_{(2)t}\nonumber\\
&&\!\!-2H_{(0)}^2e^{-2\Phi_{(0)}}\Phi_{(2)}\,,\label{eq:h2ttapp}\\
h_{(2,1)ti}\!\! & = &\!\! -H_{(0)}A_{(2,1)i}\,,\\
h_{(2)ti}\!\! & = &\!\! \frac{1}{4}H_{(0)}{}^{-1}h_{(0)t}{}^jD^{(0)}_i\partial_jH_{(0)}+\frac{1}{4}D^{(0)j}h_{(0)tj}H_{(0)}^{-1}\partial_iH_{(0)}-
\frac{1}{4}D^{(0)}_ih_{(0)tj}H_{(0)}^{-1}\partial^jH_{(0)}\nonumber\\
&&\!\!+\frac{1}{4}D^{(0)j}\left(D^{(0)}_jh_{(0)ti}-D^{(0)}_ih_{(0)tj}\right)-
\frac{1}{4}H_{(0)}^{-1}D^{(0)j}\partial_jH_{(0)}h_{(0)ti}\nonumber\\
&&\!\!+\frac{1}{4}H_{(0)}^{-1}\partial_t\partial_iH_{(0)}-\frac{1}{4}H_{(0)}^{-2}\partial_tH_{(0)}\partial_iH_{(0)}-\frac{1}{4}D^{(0)j}\partial_t\Pi_{(0)ij}+\frac{1}{4}\Pi^{jk}_{(0)}D^{(0)}_i\partial_t\Pi_{(0)jk}\nonumber\\
&&\!\!-\frac{1}{8}\Pi^{jk}_{(0)}\partial_t\Pi_{(0)jk}H_{(0)}^{-1}\partial_iH_{(0)}+\frac{1}{4}\partial_t\Phi_{(0)}\partial_i\Phi_{(0)}-\frac{1}{k}e^{2\Phi_{(0)}}H_{(0)}^{-1}\partial_t\chi_{(0)}h_{(0)ti}\nonumber\\
&&\!\!+H_{(0)}^{-1}A_{(2)t}e^{2\Phi_{(0)}}h_{(0)ti}-H_{(0)}A_{(2)i}+2\Phi_{(2)}h_{(0)ti}\,,\\
h_{(2)ij}\!\! & = &\!\! -\frac{1}{2}\mathcal{R}_{(0)ij} + \frac{1}{2}H_{(0)}^{-1}D^{(0)}_i\partial_jH_{(0)}-\frac{1}{4}H_{(0)}^{-2}\partial_iH_{(0)}\partial_jH_{(0)}+\frac{1}{4}\partial_i\Phi_{(0)}\partial_j\Phi_{(0)} \nonumber\\
&&\!\! +H_{(0)}^{-1}\Pi_{(0)ij}\left(A_{(2)t}e^{2\Phi_{(0)}}+2H_{(0)}\Phi_{(2)}-\frac{1}{4}D^{(0)k}\partial_kH_{(0)}-\frac{1}{k}e^{2\Phi_{(0)}}\partial_t\chi_{(0)}\right) \,,\\
A_{(2,1)t}\!\! &=&\!\! -2H_{(0)}e^{-2\Phi_{(0)}}\Phi_{(2,1)} \,,\\
A_{(2)t}\!\!&=&\!\! -2\Phi_{(2)}H_{(0)}e^{-2\Phi_{(0)}}+\frac{2}{3k}\partial_t\chi_{(0)}+\frac{1}{3k}h_{(0)t}{}^i\partial_i\chi_{(0)}+\frac{1}{6k^2}H_{(0)}\partial_i\chi_{(0)}\partial^i\chi_{(0)}\nonumber\\
&&\!\!+\frac{1}{12}H_{(0)}e^{-2\Phi_{(0)}}\left(\mathcal{R}_{(0)}+H_{(0)}^{-1}D^{(0)i}\partial_i H_{(0)}+\frac{1}{2}H_{(0)}^{-2}\partial_i H_{(0)}\partial^i H_{(0)}\right.\nonumber\\
&&\!\!\left.-\frac{1}{2}\partial_i\Phi_{(0)}\partial^i\Phi_{(0)}\right) \,,\\
A_{(0)i}\!\!&=&\!\!  \frac{1}{k}\partial_i\chi_{(0)} \,,
\end{eqnarray}
\begin{eqnarray}
A_{(2,1)i}\!\! & = &\!\! \frac{1}{4}H_{(0)}^{-1}\mathcal{R}_{(0)ij}h_{(0)t}{}^j+\frac{1}{4}H_{(0)}^{-1}D^{(0)j}D^{(0)}_jh_{(0)ti}+\frac{1}{4}H_{(0)}^{-2}\partial_iH_{(0)}D^{(0)}_jh_{(0)t}{}^j\nonumber\\
&&\!\!-\frac{1}{4}H_{(0)}^{-2}\partial^jH_{(0)}\left(D^{(0)}_ih_{(0)tj}+D^{(0)}_jh_{(0)ti}\right)-\frac{1}{2}H_{(0)}^{-1}D^{(0)}_jh_{(0)t}{}^j\partial_i\Phi_{(0)}\nonumber\\
&&\!\!+\frac{1}{2}H_{(0)}^{-1}\partial^j\Phi_{(0)}\left(D^{(0)}_ih_{(0)tj}+D^{(0)}_jh_{(0)ti}\right)-\frac{5}{4}H_{(0)}^{-1}h_{(0)t}{}^j\partial_i\Phi_{(0)}\partial_j\Phi_{(0)}\nonumber\\
&&\!\!-\frac{1}{4}H_{(0)}^{-1}D^{(0)j}\partial_t\Pi_{(0)ij}+\frac{1}{8}H_{(0)}^{-1}\Pi_{(0)}^{jk}D^{(0)}_i\partial_t\Pi_{(0)jk}+\frac{1}{4}H_{(0)}^{-2}\partial^jH_{(0)}\partial_t\Pi_{(0)ij}\nonumber\\
&&\!\!-\frac{1}{8}H_{(0)}^{-2}\partial_iH_{(0)}\Pi_{(0)}^{jk}\partial_t\Pi_{(0)jk}+\frac{1}{4}H_{(0)}^{-1}\partial_i\Phi_{(0)}\Pi_{(0)}^{jk}\partial_t\Pi_{(0)jk}\nonumber\\
&&\!\!-\frac{1}{2}H_{(0)}^{-1}\partial^j\Phi_{(0)}\partial_t\Pi_{(0)ij}+\frac{5}{4}H_{(0)}^{-1}\partial_t\Phi_{(0)}\partial_i\Phi_{(0)}\,,
\end{eqnarray}
\begin{eqnarray}
\Phi_{(2,1)}\!\! &=&\!\! \frac{1}{12}\mathcal{R}_{(0)}+\frac{1}{12}D^{(0)i}\partial_i\log H_{(0)}-\frac{1}{2}D^{(0)i}\partial_i\Phi_{(0)}-\frac{13}{24}\partial_i\Phi_{(0)}\partial^i\Phi_{(0)}\nonumber\\
&&\!\!+\frac{4}{3k^2}e^{2\Phi_{(0)}}\left(-\partial_i\chi_{(0)}\partial^i\chi_{(0)}+2kH_{(0)}^{-1}\partial_t\chi_{(0)}-2kH_{(0)}^{-1}h_{(0)t}{}^i\partial_i\chi_{(0)}\right)\,,\\
\phi_{(2)} & = & \frac{1}{4}D^{(0)i}\partial_i\Phi_{(0)} + \frac{1}{4}H_{(0)}^{-1}\partial_iH_{(0)}\partial^i\Phi_{(0)} \nonumber\\
&&\!\! + \frac{1}{k}H_{(0)}^{-1}e^{2\Phi_{(0)}}\left(\frac{1}{k}H_{(0)}\partial_i\chi_{(0)} \partial^i\chi_{(0)} + 2\partial_i\chi_{(0)} h_{(0)t}{}^i - 2\partial_t\chi_{(0)}\right) \,,\\
\chi_{(2)} & = &  \frac{k}{2}H_{(0)}^{-1}\left(\frac{1}{4}\Pi^{ij}_{(0)}\partial_t\Pi_{(0)ij}- \frac{1}{2}D^{(0)i}h_{(0)ti} - h_{(0)t}{}^i\partial_i\Phi_{(0)} + \partial_t\Phi_{(0)}\right)\,,\label{eq:phi2}
\end{eqnarray}
We have not listed coefficients of the form $a_{(4,m)}$ for some field $a$ even though they can be computed from the reduction, the expressions are typically half a page and so we will not write them. Further we did not write coefficients that depend explicitly on the reduction of $\hat t_{\hat a\hat b}$ as these can be considered `arbitrary' from a 4-dimensional point of view\footnote{The coefficients that depend explicitly on $\hat t_{\hat a\hat b}$ are $\Phi_{(2)}$ ($\hat t_{uu}$), $A_{(2)i}$ ($\hat t_{iu}$), $A_{(4)t}$ ($\hat t_{tu}$), $h_{(6)tt}$ ($\hat t_{tt}$), $h_{(4)ti}$ ($\hat t_{ti}$) and $h_{(4)ij}$ ($\hat t_{ij}$).}. We put arbitrary in quotation marks because these coefficients are constrained by the reduced version of equations \eqref{eq:tracet} and \eqref{eq:divt}.

\bibliographystyle{utphysmodb}

\bibliography{LifshitzHR}

\providecommand{\href}[2]{#2}\begingroup\raggedright\begin{thebibliography}{10}

\bibitem{Hartnoll:2009sz}
S.~A. Hartnoll,  {\em {Lectures on holographic methods for condensed matter
  physics}}, Class.Quant.Grav. {\bf 26} (2009) 224002
[\href{http://www.arXiv.org/abs/0903.3246}{{\tt 0903.3246}}].

\bibitem{McGreevy:2009xe}
J.~McGreevy,  {\em {Holographic duality with a view toward many-body physics}},
  Adv.High Energy Phys. {\bf 2010} (2010) 723105
[\href{http://www.arXiv.org/abs/0909.0518}{{\tt 0909.0518}}].

\bibitem{Sachdev:2010ch}
S.~Sachdev,  {\em {Condensed Matter and AdS/CFT}},
\href{http://www.arXiv.org/abs/1002.2947}{{\tt 1002.2947}}.

\bibitem{Koroteev:2007yp}
P.~Koroteev and M.~Libanov,  {\em {On Existence of Self-Tuning Solutions in
  Static Braneworlds without Singularities}}, JHEP {\bf 0802} (2008) 104
[\href{http://www.arXiv.org/abs/0712.1136}{{\tt 0712.1136}}].

\bibitem{Kachru:2008yh}
S.~Kachru, X.~Liu and M.~Mulligan,  {\em {Gravity Duals of Lifshitz-like Fixed
  Points}}, Phys.Rev. {\bf D78} (2008) 106005
[\href{http://www.arXiv.org/abs/0808.1725}{{\tt 0808.1725}}].

\bibitem{Son:2008ye}
D.~Son,  {\em {Toward an AdS/cold atoms correspondence: A Geometric realization
  of the Schrodinger symmetry}}, Phys.Rev. {\bf D78} (2008) 046003
[\href{http://www.arXiv.org/abs/0804.3972}{{\tt 0804.3972}}].

\bibitem{Balasubramanian:2008dm}
K.~Balasubramanian and J.~McGreevy,  {\em {Gravity duals for non-relativistic
  CFTs}}, Phys.Rev.Lett. {\bf 101} (2008) 061601
[\href{http://www.arXiv.org/abs/0804.4053}{{\tt 0804.4053}}].

\bibitem{Copsey:2010ya}
K.~Copsey and R.~Mann,  {\em {Pathologies in Asymptotically Lifshitz
  Spacetimes}}, JHEP {\bf 1103} (2011) 039
[\href{http://www.arXiv.org/abs/1011.3502}{{\tt 1011.3502}}].

\bibitem{Horowitz:2011gh}
G.~T. Horowitz and B.~Way,  {\em {Lifshitz Singularities}},
\href{http://www.arXiv.org/abs/1111.1243}{{\tt 1111.1243}}.

\bibitem{Ross:2011gu}
S.~F. Ross,  {\em {Holography for asymptotically locally Lifshitz spacetimes}},
  Class.Quant.Grav. {\bf 28} (2011) 215019
[\href{http://www.arXiv.org/abs/1107.4451}{{\tt 1107.4451}}].

\bibitem{Ross:2009ar}
S.~F. Ross and O.~Saremi,  {\em {Holographic stress tensor for non-relativistic
  theories}}, JHEP {\bf 0909} (2009) 009
[\href{http://www.arXiv.org/abs/0907.1846}{{\tt 0907.1846}}].

\bibitem{Baggio:2011cp}
M.~Baggio, J.~de~Boer and K.~Holsheimer,  {\em {Hamilton-Jacobi Renormalization
  for Lifshitz Spacetime}},
\href{http://www.arXiv.org/abs/1107.5562}{{\tt 1107.5562}}.

\bibitem{Mann:2011hg}
R.~B. Mann and R.~McNees,  {\em {Holographic Renormalization for Asymptotically
  Lifshitz Spacetimes}}, JHEP {\bf 1110} (2011) 129
[\href{http://www.arXiv.org/abs/1107.5792}{{\tt 1107.5792}}].

\bibitem{Griffin:2011xs}
T.~Griffin, P.~Horava and C.~M. Melby-Thompson,  {\em {Conformal Lifshitz
  Gravity from Holography}},
\href{http://www.arXiv.org/abs/1112.5660}{{\tt 1112.5660}}.

\bibitem{Baggio:2011ha}
M.~Baggio, J.~de~Boer and K.~Holsheimer,  {\em {Anomalous Breaking of
  Anisotropic Scaling Symmetry in the Quantum Lifshitz Model}},
\href{http://www.arXiv.org/abs/1112.6416}{{\tt 1112.6416}}.

\bibitem{Balasubramanian:2010uk}
K.~Balasubramanian and K.~Narayan,  {\em {Lifshitz spacetimes from AdS null and
  cosmological solutions}}, JHEP {\bf 1008} (2010) 014
[\href{http://www.arXiv.org/abs/1005.3291}{{\tt 1005.3291}}].

\bibitem{Donos:2010tu}
A.~Donos and J.~P. Gauntlett,  {\em {Lifshitz Solutions of D=10 and D=11
  supergravity}}, JHEP {\bf 1012} (2010) 002
[\href{http://www.arXiv.org/abs/1008.2062}{{\tt 1008.2062}}].

\bibitem{Cassani:2011sv}
D.~Cassani and A.~F. Faedo,  {\em {Constructing Lifshitz solutions from AdS}},
  JHEP {\bf 1105} (2011) 013
[\href{http://www.arXiv.org/abs/1102.5344}{{\tt 1102.5344}}].

\bibitem{Halmagyi:2011xh}
N.~Halmagyi, M.~Petrini and A.~Zaffaroni,  {\em {Non-Relativistic Solutions of
  N=2 Gauged Supergravity}}, JHEP {\bf 1108} (2011) 041
[\href{http://www.arXiv.org/abs/1102.5740}{{\tt 1102.5740}}].

\bibitem{Gregory:2010gx}
R.~Gregory, S.~L. Parameswaran, G.~Tasinato and I.~Zavala,  {\em {Lifshitz
  solutions in supergravity and string theory}}, JHEP {\bf 1012} (2010) 047
[\href{http://www.arXiv.org/abs/1009.3445}{{\tt 1009.3445}}].

\bibitem{Chemissany:2011mb}
W.~Chemissany and J.~Hartong,  {\em {From D3-Branes to Lifshitz Space-Times}},
  Class.Quant.Grav. {\bf 28} (2011) 195011
[\href{http://www.arXiv.org/abs/1105.0612}{{\tt 1105.0612}}].

\bibitem{Adam:2009gq}
I.~Adam, I.~V. Melnikov and S.~Theisen,  {\em {A Non-Relativistic Weyl
  Anomaly}}, JHEP {\bf 0909} (2009) 130
[\href{http://www.arXiv.org/abs/0907.2156}{{\tt 0907.2156}}].

\bibitem{Gomes:2011di}
P.~R. Gomes and M.~Gomes,  {\em {On Ward Identities in Lifshitz-like Field
  Theories}},
\href{http://www.arXiv.org/abs/1112.3887}{{\tt 1112.3887}}.

\bibitem{Henningson:1998gx}
M.~Henningson and K.~Skenderis,  {\em {The Holographic Weyl anomaly}}, JHEP
  {\bf 9807} (1998) 023
[\href{http://www.arXiv.org/abs/hep-th/9806087}{{\tt hep-th/9806087}}].

\bibitem{Papadimitriou:2011qb}
I.~Papadimitriou,  {\em {Holographic Renormalization of general dilaton-axion
  gravity}}, JHEP {\bf 1108} (2011) 119
[\href{http://www.arXiv.org/abs/1106.4826}{{\tt 1106.4826}}].

\bibitem{deHaro:2000xn}
S.~de~Haro, S.~N. Solodukhin and K.~Skenderis,  {\em {Holographic
  reconstruction of space-time and renormalization in the AdS / CFT
  correspondence}}, Commun.Math.Phys. {\bf 217} (2001) 595--622
[\href{http://www.arXiv.org/abs/hep-th/0002230}{{\tt hep-th/0002230}}].

\bibitem{Papadimitriou:2005ii}
I.~Papadimitriou and K.~Skenderis,  {\em {Thermodynamics of asymptotically
  locally AdS spacetimes}}, JHEP {\bf 0508} (2005) 004
[\href{http://www.arXiv.org/abs/hep-th/0505190}{{\tt hep-th/0505190}}].

\bibitem{FeffermanGraham}
C.~Fefferman and C.~R. Graham,  {\em {Conformal Invariants}}, Elie Cartan et
  les Math\'ematiques d'aujourd'hui (Asterisque) {\bf 1103} (1985) 95.

\bibitem{Graham:1999jg}
C.~R. Graham,  {\em {Volume and area renormalizations for conformally compact
  Einstein metrics}},
\href{http://www.arXiv.org/abs/math/9909042}{{\tt math/9909042}}.

\bibitem{Taylor:2008tg}
M.~Taylor,  {\em {Non-relativistic holography}},
\href{http://www.arXiv.org/abs/0812.0530}{{\tt 0812.0530}}.

\bibitem{Costa:2010cn}
R.~Caldeira~Costa and M.~Taylor,  {\em {Holography for chiral scale-invariant
  models}}, JHEP {\bf 1102} (2011) 082
[\href{http://www.arXiv.org/abs/1010.4800}{{\tt 1010.4800}}].

\bibitem{Kanitscheider:2008kd}
I.~Kanitscheider, K.~Skenderis and M.~Taylor,  {\em {Precision holography for
  non-conformal branes}}, JHEP {\bf 0809} (2008) 094
[\href{http://www.arXiv.org/abs/0807.3324}{{\tt 0807.3324}}].

\bibitem{Gouteraux:2011qh}
B.~Gouteraux, J.~Smolic, M.~Smolic, K.~Skenderis and M.~Taylor,  {\em
  {Holography for Einstein-Maxwell-dilaton theories from generalized
  dimensional reduction}}, JHEP {\bf 1201} (2012) 089
[\href{http://www.arXiv.org/abs/1110.2320}{{\tt 1110.2320}}].

\bibitem{Hartong:2012sw}
J.~Hartong and B.~Rollier,  {\em {Asymptotically Schr\"odinger Space-Times}},
\href{http://www.arXiv.org/abs/1202.1433}{{\tt 1202.1433}}.

\bibitem{Penrose:1986ca}
R.~Penrose and W.~Rindler, {\em Spinors and Space-Time. Vol. 2: Spinor and
  Twistor Methods in Space-Time Geometry}.
\newblock Cambridge University Press, 1986.

\bibitem{Brown:1986nw}
J.~D. Brown and M.~Henneaux,  {\em {Central Charges in the Canonical
  Realization of Asymptotic Symmetries: An Example from Three-Dimensional
  Gravity}}, Commun.Math.Phys. {\bf 104} (1986)
207--226.

\bibitem{Hartong:2010ec}
J.~Hartong and B.~Rollier,  {\em {Asymptotically Schroedinger Space-Times: TsT
  Transformations and Thermodynamics}}, JHEP {\bf 1101} (2011) 084
[\href{http://www.arXiv.org/abs/1009.4997}{{\tt 1009.4997}}].

\bibitem{Horava:2009vy}
P.~Horava and C.~M. Melby-Thompson,  {\em {Anisotropic Conformal Infinity}},
  Gen. Rel. Grav. {\bf 43} (2011) 1391--1400
[\href{http://www.arXiv.org/abs/0909.3841}{{\tt 0909.3841}}].

\bibitem{Peeters:2006kp}
K.~Peeters,  {\em {A Field-theory motivated approach to symbolic computer
  algebra}}, Comput.Phys.Commun. {\bf 176} (2007) 550--558
[\href{http://www.arXiv.org/abs/cs/0608005}{{\tt cs/0608005}}].

\bibitem{Peeters:2007wn}
K.~Peeters,  {\em {Introducing Cadabra: A Symbolic computer algebra system for
  field theory problems}},
\href{http://www.arXiv.org/abs/hep-th/0701238}{{\tt hep-th/0701238}}.

\bibitem{Horava:2008ih}
P.~Horava,  {\em {Membranes at Quantum Criticality}}, JHEP {\bf 0903} (2009)
  020
[\href{http://www.arXiv.org/abs/0812.4287}{{\tt 0812.4287}}].

\bibitem{Balasubramanian:2011ua}
K.~Balasubramanian and J.~McGreevy,  {\em {String theory duals of
  Lifshitz-Chern-Simons gauge theories}},
\href{http://www.arXiv.org/abs/1111.0634}{{\tt 1111.0634}}.

\end{thebibliography}\endgroup

\end{document}